\documentclass[multphys,vecphys]{svmult}

\usepackage{epsf,epsfig}

\usepackage{amsmath}
\usepackage{amsfonts}
\usepackage{amssymb}
\usepackage{makeidx}         % allows index generation
\usepackage{graphicx}        % standard LaTeX graphics tool
\usepackage{multicol}        % used for the two-column index
\usepackage[bottom]{footmisc}% places footnotes at page bottom
\makeindex             % used for the subject index
\begin{document}

\title*{Spacetime metric from local and linear electrodynamics: a new
  axiomatic scheme} \titlerunning{Spacetime metric from local and
  linear electrodynamics}
\author{Friedrich W. Hehl\inst{1}\and
Yuri N. Obukhov\inst{2}}
\institute{Inst. Theor. Physics, University of Cologne, 
50923 K\"oln, Germany\\ Dept. Physics Astron., Univ. of
Missouri-Columbia, Columbia, MO 65211, USA
\texttt{hehl@thp.uni-koeln.de}
\and Inst. Theor. Physics, University of Cologne, 
50923 K\"oln, Germany\\ Dept. Theor. Physics, Moscow State University
 117234 Moscow, Russia
\texttt{yo@thp.uni-koeln.de}}
\maketitle 
{Invited lecture, 339th WE Heraeus Seminar on Special
  Relativity\\Potsdam, Germany, 13-18 February 2005}
\begin{abstract}
  We consider {\it spacetime\/} to be a 4-dimensional differentiable
  manifold that can be split locally into time and space. No metric,
  no linear connection are assumed. {\it Matter\/} is described by
  classical fields/fluids. We distinguish electrically charged from
  neutral matter. {\it Electric charge\/} and {\it magnetic flux\/}
  are postulated to be conserved. As a consequence, the inhomogeneous
  and the homogeneous Maxwell equations emerge expressed in terms of
  the excitation $H=({\cal H}, {\cal D})$ and the field strength
  ${F}=(E,B)$, respectively. $H$ and $F$ are assumed to fulfill a {\it
    local and linear\/} ``spacetime relation'' with 36 constitutive
  functions.  The propagation of electromagnetic waves is considered
  under such circumstances in the geometric optics limit. We forbid
  {\it birefringence\/} in vacuum and find the light cone including
  its Lorentzian signature. Thus the conformally invariant part of the
  metric is recovered. If one sets a scale, one finds the
  pseudo-Riemannian metric of spacetime.

%\hfill  {\it file axiomatics11.tex, 2005-08-04}
\end{abstract}

% \begin{footnotesize}
% \noindent Summary\vspace{-5pt}
% \begin{enumerate}
% \item Introduction
% \item Spacetime
% \item Matter --- electrically charged and neutral
% \item Electric charge conservation
% \item Charge active: excitation
% \item Charge passive: field strength
% \item Magnetic flux conservation
% \item Premetric electrodynamics
% \item The excitation is local and linear in the field strength
% \item Propagation of electromagnetic rays (``light'')
% \item No birefringence in vacuum and the light cone
% \item Dilaton, metric, axion
% \item Setting the scale
% \item Discussion
% \end{enumerate}  \vspace{-6pt}
% \noindent Acknowledgments
% 
% \noindent References
% \end{footnotesize}

%%%%%%%%%%%%%%%%%%%%%%%%%%%%%%%%%%%%%%%%%%%%%%%%%%%%%%%%%%%%%%%%%%%
\section{Introduction}
%%%%%%%%%%%%%%%%%%%%%%%%%%%%%%%%%%%%%%%%%%%%%%%%%%%%%%%%%%%%%%%%%%%

The neutrinos, in the standard model of elementary particle physics,
are assumed to be massless. By the discovery of the neutrino
oscillations, this assumption became invalidated. The neutrinos are
massive, even though they carry, as compared to the electron, only
very small masses. Then the {\it photon\/} is left as the only known
massless and free elementary particle. The gluons do not qualify in
this context since they are confined and cannot exist as free
particles under normal circumstances.

Consequently, the photon is the only particle that is directly related
to the light cone $g_{ij}dx^i\otimes dx^j=0$ and that can be used for
an operational definition of the light cone; here $g_{ij}$ is the
metric of spacetime, $dx^i$ a coordinate differential, and
$i,j=0,1,2,3$. We are back --- as the name light cone suggests anyway
--- to an electromagnetic view of the light cone.  Speaking in the
framework of classical optics, the {\it light ray\/} would then be the
elementary object with the help of which one can span the light cone.
We take ``light ray'' as synonymous for radar signals, laser beams, or
electromagnetic rays of other wavelengths. It is understood that
classical optics is a limiting case, for short wavelengths, of
classical Maxwell-Lorentz electrodynamics.

In other words, if we assume the framework of Maxwell-Lorentz
electrodynamics, we can derive, in the geometric optics limit, light
rays and thus the light cone, see Perlick \cite{Perlick00} and the
literature given there. However, the formalism of Maxwell-Lorentz
electrodynamics is interwoven with the Riemannian metric $g_{ij}$ in a
nontrivial way.  Accordingly, in the way sketched, one can never hope
to find a real derivation of the light cone.

Therefore, we start from the {\it premetric\/} form of
electrodynamics, that is, a metric of spacetime is not assumed.
Nevertheless, we can derive the generally covariant Maxwell equations,
expressed in terms of the excitation $H=({\cal H}, {\cal D})$ and the
field strength ${F}=(E,B)$, from the conservation laws of electric
charge and magnetic flux. We assume a local and linear spacetime
relation between $H$ and $F$. Then we can solve the Maxwell equations.
In particular, we can study the propagation of electromagnetic waves,
and we can consider the geometrical optics limit. In this way, we
derive the light rays that are spanning the light cone. In general, we
find a {\it quartic\/} wave covector surface (similar as in a crystal)
that only reduces to the pseudo-Riemannian light cone of general
relativity if we forbid {\it birefringence\/} (double refraction) in
vacuum. Hence, in the framework of premetric electrodynamics, the
local and linear spacetime relation, together with a ban on
birefringence in vacuum, yields the pseudo-Riemannian light cone of
general relativity. Accordingly, the geometrical structure of a
Riemannian spacetime is derived from purely electromagnetic data. We
consider that as our contribution to the Einstein year 2005, and we
hope that going beyond the geometrical optics limit will yield better
insight into the geometry of spacetime.

The axiomatic scheme that we are going to present here is already
contained in our book \cite{Birkbook} where also references to earlier
work and more details can be found. In the meantime we learnt from the
literature that appeared since 2003 (see, e.g., Delphenich
\cite{Delphenich1,Delphenich2}, Itin \cite{Itin}, Kaiser
\cite{Kaiser}, Kiehn \cite{Kiehn04}, and Lindell \& Sihvola
\cite{Lindell,LindSihv2004a}) and improved our derivation of the light
cone, simplified it, made it more transparent (see, e.g.,
\cite{ourmonopole,PostCon,Laemmerzahl04,skewon,measure}).  The
formalism and the conventions we take from \cite{Birkbook}.

%%%%%%%%%%%%%%%%%%%%%%%%%%%%%%%%%%%%%%%%%%%%%%%%%%%%%%%%%%%%%%%%%%%
\section{Spacetime}
%%%%%%%%%%%%%%%%%%%%%%%%%%%%%%%%%%%%%%%%%%%%%%%%%%%%%%%%%%%%%%%%%%%

In our approach, we start from a 4-dimensional spacetime manifold that
is just a continuum which can be decomposed locally in (1-dimensional)
time and (3-dimensional) space.  It carries no metric and it carries
no (linear or affine) connection.  As such it is inhomogeneous. It
doesn't make sense to assume that a vector field is constant in this
continuum. Only the constancy of a scalar field is uniquely defined.
Also a measurement of temporal or spatial intervals is still not
defined since a metric is not yet available.

\begin{figure}
\centering
\includegraphics[height=4cm]{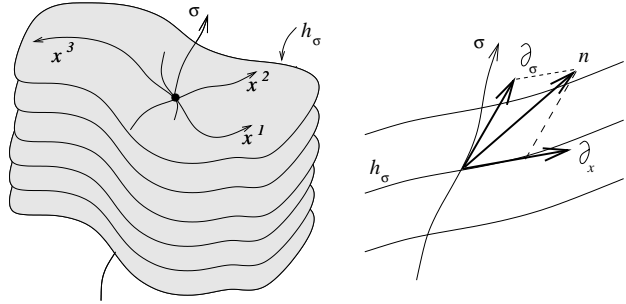}
%\picplace{5cm}{2cm} % Give the correct figure height and width in cm
%
\caption{Local spacetime foliation, see \cite{Birkbook}.}
\label{fig:1}       % Give a unique label
\end{figure}

In technical terms, the spacetime is a 4-dimen\-sional connected,
Hausdorff, paracompact, and oriented differential manifold.  On such a
manifold, we assume the existence of a foliation: The spacetime can be
decomposed locally into three-dimensional folios labeled consecutively
by a monotonically increasing ``prototime'' parameter $\sigma$, see
Fig.~\ref{fig:1}.
A vector field $n$, transverse to the foliation, is normalized by
$n\rfloor d\sigma={\cal L}_n\sigma=1$. Accordingly, we find for the
dimensions $[n]=[\sigma]^{-1}=t^{-1}$, where $t$ denotes the dimension
of time.

We can decompose any exterior form $\Psi$ in ``time'' and ``space''
pieces. The part {\it longitudinal\/} to the vector $n$ reads
\begin{equation}\label{longi}
^{\bot}\Psi := d\sigma\wedge\Psi_{\bot}\,,\qquad 
\Psi_{\bot}:= {n}\rfloor\Psi\,,
\end{equation} 
the part {\it transversal\/} to the vector $n$
\begin{equation}\label{trans}
  \underline{\Psi}:=(1\;-\;^{\bot})\Psi=
  {n}\rfloor(d\sigma\wedge\Psi)\, , \quad{n}\rfloor\,
  \underline{\Psi}\equiv 0\, .
\end{equation}
Putting these two parts together, we have the space-time decomposition
\begin{equation}\label{decomppform}
  \Psi=\,^{\bot}\Psi+ \underline{\Psi}= d\sigma\wedge\Psi_{\bot}+
  \underline{\Psi}\,,
\end{equation} 
with the absolute dimensions $[\Psi_{\bot}]=[\Psi]\,t^{-1}$ and
$[\underline{\Psi}]=[\Psi]$. 

The 3-dimensional exterior derivative is defined by $\underline{d}:=
n\rfloor(d\sigma\wedge d)$. We can use the notion of the {\it Lie
  derivative} of a $p$-form $\Psi$ along a vector field $\xi$, i.e.,
${\cal L}_{\xi}\Psi := \xi \rfloor d\Psi + d(\xi\rfloor\Psi)$, and can
introduce the derivative of a transversal field $\underline{\Psi}$
with respect to prototime as
\begin{equation}\label{dot}
\dot{\underline{\Psi}}:={\cal L}_n \underline{\Psi}\,.
\end{equation} 

%%%%%%%%%%%%%%%%%%%%%%%%%%%%%%%%%%%%%%%%%%%%%%%%%%%%%%%%%%%%%%%%%%%
\section{Matter --- electrically charged and neutral}\label{3}
%%%%%%%%%%%%%%%%%%%%%%%%%%%%%%%%%%%%%%%%%%%%%%%%%%%%%%%%%%%%%%%%%%%

We assume that spacetime is ``populated'' with classical matter,
either described by fields and/or by fluids. In between the
agglomerations of matter, there may also exist vacuum.

Matter is divided into electrically charged and neutral matter.
Turning to the physics of the former, we assume that on the spatial
folios of the manifold we can determine an electric charge $Q$ as a
3-dimensional integral over a charge density and a magnetic flux
$\Phi$ as a 2-dimensional integral over a flux density.

This is at the bottom of classical electrodynamics: Spacetime is
filled with matter that is characterized by charge $Q$ and by magnetic
flux $\Phi$. For neutral matter both vanish. The absolute dimension of
charge will be denoted by $q$, that of magnetic flux by $\phi$ =
[action/charge] = $h/q$, with $h$ as the dimension of action.

%%%%%%%%%%%%%%%%%%%%%%%%%%%%%%%%%%%%%%%%%%%%%%%%%%%%%%%%%%%%%%%%%%%
\section{Electric charge conservation}
%%%%%%%%%%%%%%%%%%%%%%%%%%%%%%%%%%%%%%%%%%%%%%%%%%%%%%%%%%%%%%%%%%%

One can catch single electrons and single protons in traps and can
{\it count\/} them individually. Thus, the electric charge
conservation is a fundamental law of nature, valid in macro- as well
as in micro-physics.\footnote{L\"ammerzahl, Macias, and M\"uller
  \cite{chargenoncons} proposed an extension of Maxwell's equations
  that violates electric charge conservation.  Such a model can be used
  as a test theory for experiments that check the validity of charge
  conservation, and it allows to give a numerical bound. }
% Elementary charge $e$ (or, in the case of quarks, in
% $1/3$th of it) can be counted in principle. 
Accordingly, it is justified to introduce the absolute dimension of
charge $q$ as a new and independent concept.

Let us define, in 4-dimensional spacetime, the electrical current
3-form $J$, with dimension $[J]=q$. Its integral over an arbitrary
3-dimensional spacetime domain yields the total charge contained
therein: $Q=\int_{\Omega_3}J$.
%  The total charge can be determined by counting the particles
%  carrying an elementary charge. 
Accordingly, the local form of charge conservation (Axiom 1) reads:
\begin{equation}
d\,J = 0\,.\label{dJ1}
\end{equation}
This law is metric-independent since it is based on a {\it counting\/}
procedure for the elementary charges. Using a foliation of spacetime,
we can decompose the current $J$ into the 2-form of the electric
current density $j$ and the 3-form $\rho$ of the electric charge
density:
\begin{equation}\label{decompJ}
J=-j\wedge d\sigma +\rho\,.
\end{equation}
Then (\ref{dJ1}) can be rewritten as the continuity equation:
\begin{equation}
\dot{\rho} + \underline{d}\,j = 0. \label{dJ2}
\end{equation}
Both versions of charge conservation, eqs.(\ref{dJ1}) and \eqref{dJ2},
can also be formulated in an integral form.

%%%%%%%%%%%%%%%%%%%%%%%%%%%%%%%%%%%%%%%%%%%%%%%%%%%%%%%%%%%%%%%%%%%
\section{Charge active: excitation}
%%%%%%%%%%%%%%%%%%%%%%%%%%%%%%%%%%%%%%%%%%%%%%%%%%%%%%%%%%%%%%%%%%%

Electric charge was postulated to be conserved in all regions of
spacetime. If spacetime is topologically sufficiently trivial, we
find, as consequence of \eqref{dJ1}, that $J$ has to be exact:
\begin{equation}
J = d\,H\,.\label{JdH}
\end{equation}
This is the inhomogeneous Maxwell equation in its premetric form. The
{\it electromagnetic excitation} 2-form $H$, with $[H]=[J]=q$, is
measurable with the help of ideal conductors and superconductors
and thus has a direct {\it operational\/} significance.

By decomposing $H$ into time and space, we obtain the electric
excitation 2-form ${{\cal D}}$ (historical name: ``dielectric
displacement") and the magnetic excitation 1-form ${{\cal H}}$
(``magnetic field"):
\begin{equation}
H = -{\cal H}\wedge d\sigma + {\cal D}\,.\label{HHD}
\end{equation}
Substituting (\ref{HHD}) into (\ref{JdH}), we recover the pair of the 
3-dimensional inhomogeneous Maxwell equations
\begin{equation}\label{axiom1}
  d\,H= J\>\,\,\begin{cases}
\hspace{30pt}\underline{d}\,{\cal D}\>=\rho\,,\\ 
-\,\dot{\cal D}+\underline{d}\,{\cal H}\>=j\,.
\end{cases}
\end{equation}

%%%%%%%%%%%%%%%%%%%%%%%%%%%%%%%%%%%%%%%%%%%%%%%%%%%%%%%%%%%%%%%%%%%
\section{Charge passive: field strength}
%%%%%%%%%%%%%%%%%%%%%%%%%%%%%%%%%%%%%%%%%%%%%%%%%%%%%%%%%%%%%%%%%%%

With the derivation of the inhomogeneous Maxwell equations the
information contained in Axiom 1 is exhausted. As is evident from the
Coulomb-Gauss law $\underline{d}\,{\cal D}=\rho$, it is the active
character of $\rho$ that plays a role in this inhomogeneous Maxwell
equation: The charge density $\rho$ is the source of ${\cal D}$ (and,
analogously, the current density $j$ that of ${\cal H}$).

Since we search for new input, it is near at hand to turn to the {\it
  passive\/} character of charge, that is, to wonder what happens when
a test charge is put in an electromagnetic field. In the purely
electric case with a test charge $e$, we have
\begin{equation}\label{Coul}
{F}_a\sim e\,{E_a}\,,\
\end{equation}
with ${F}_a$ and $E_a$ as components of the covectors of force and
electric field strength, respectively.  The simplest relativistic
generalization for defining the electromagnetic field is then of the
type
\begin{equation}\label{fieldansatz1}
{\rm force\>\> density}\sim {\rm field\>\>strength}
  \times{\rm charge\>\> current\>\> density}\,.
\end{equation} 

\begin{figure}
\centering
\includegraphics[height=6cm]{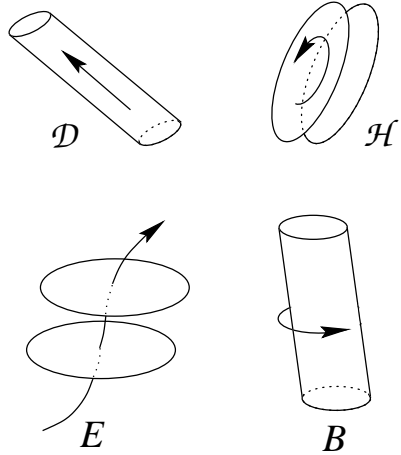}
%\picplace{5cm}{2cm} % Give the correct figure height and width in cm
%
\caption{Faraday-Schouten pictograms of the electromagnetic field, 
  see \cite{Birkbook}. The electric excitation $\cal D$ is a twisted
  2-form, the magnetic excitation $\cal H$ a twisted 1-form. The
  electric field strength $E$ is a 1-form and the magnetic field
  strength $B$ a 2-form, both without twist.}
\label{fig:3}       % Give a unique label
\end{figure}

Accordingly, with the force density covector (or 1-form) $f_\alpha$,
we can formulate Axiom 2 as 
\begin{equation}\label{Lorentz}
  f_\alpha =(e_\alpha\rfloor F)\wedge J\,.
\end{equation}
Here $e_\alpha$ is a local frame, with $\alpha=0,1,2,3$. Axiom 2
provides an operational definition of the electromagnetic field
strength 2-form $F$, the absolute dimension of which turns out to be
$[F]=h/q $.  Its $1+3$ decomposition
\begin{equation}\label{Fdecomp}
F = E\wedge d\sigma + B\,,
\end{equation}
introduces the electric field strength 1-form $E$ and the magnetic
field strength 2-form $B$, see Fig.\ref{fig:3}. If we substitute
\eqref{Fdecomp} and \eqref{decompJ} into \eqref{Lorentz}, we recover,
for $\alpha=1,2,3$, the Lorentz force density.

%%%%%%%%%%%%%%%%%%%%%%%%%%%%%%%%%%%%%%%%%%%%%%%%%%%%%%%%%%%%%%%%%%%
\section{Magnetic flux conservation}
%%%%%%%%%%%%%%%%%%%%%%%%%%%%%%%%%%%%%%%%%%%%%%%%%%%%%%%%%%%%%%%%%%%

The field strength $F$, as a 2-form, can be integrated over a
2-dimensional area $\Omega_2$ in 4-dimensional spacetime. This yields
the total magnetic flux $\Phi$ piercing through this area:
$\Phi=\int_{\Omega_2}F$. In close analogy to electric charge
conservation, we assume that also the flux is conserved. Then, in local
form, magnetic flux conservation (Axiom 3) reads\footnote{One can
  give up magnetic flux conservation by introducing magnetic monopoles
  according to $dF=J_{\rm magn}$. In premetric electrodynamics this
  has been done by Edelen \cite{edelen}, Kaiser \cite{Kaiser}, and by
  us \cite{ourmonopole}.  However, then one has to change Axiom 2,
  too, and the Lorentz force density picks up an additional term
  $-(e_\alpha\rfloor H)\wedge J_{\rm magn}$. This destroys Axiom 2 as
  an operational procedure for defining $F$. Moreover, magnetic
  charges have never been found.}
\begin{equation}\label{axiom2}
  d\,F=0\>\,\,\begin{cases}
    \qquad \underline{d}\,B\>=0\,,\\ 
    \dot{B}+\underline{d}\,E\>=0\,.\end{cases}
\end{equation}
The Faraday induction law and the sourcelessness of $B$ are the two
consequences of $dF=0$. In this sense, Axiom 3 has a firm experimental
underpinning.
% Note that the induction law has the form of a continuity
% equation for the ``charge density'' $B$. 

% \begin{figure}
% \centering
% \includegraphics[height=4cm]{Birk/B04flux.eps}
% \caption{Different 2D periods of the magnetic flux integral}
% \label{fig:2}       % Give a unique label
% \end{figure}

%%%%%%%%%%%%%%%%%%%%%%%%%%%%%%%%%%%%%%%%%%%%%%%%%%%%%%%%%%%%%%%%%%%
\section{Premetric electrodynamics}
%%%%%%%%%%%%%%%%%%%%%%%%%%%%%%%%%%%%%%%%%%%%%%%%%%%%%%%%%%%%%%%%%%%

...is meant to be the ``naked'' or ``featureless'' spacetime manifold,
without metric and without connection, together with the Maxwell
equations $dH=J,\,dF=0$, the Lorentz force formula, and the
electromagnetic energy-momentum current to be discussed below, see
\eqref{simax}. We stress that the Poincar\'e group and special
relativity have nothing to do with the foundations of electrodynamics
as understood here in the sense of the decisive importance of the
underlying {\it generally covariant\/} conservation laws of charge
(Axiom 1) and flux (Axiom 3).  Historically, special relativity
emerged in the context of an analysis of the electrodynamics of moving
bodies \cite{Einstein05,Dover1952}, but within the last 100 years
  classical electrodynamics had a development of its own and its
  structure is now much better understood than it was 100 years ago.
  {\it Diffeomorphism invariance\/} was recognized to be of
  overwhelming importance. Poincar\'e invariance turned out to play a
  secondary role only.

Of course, premetric electrodynamics so far does not represent a
complete physical theory. The excitation $H$ does not yet communicate
with the field strength $F$. Only by specifying a ``spacetime''
relation between $H$ and $F$ (the constitutive law of the spacetime
manifold), only thereby we recover --- under suitable conditions ---
our normal Riemannian or Minkowskian spacetime which we seem to live
in. In this sense, a realistic spacetime --- and thus an appropriate
{\it geometry\/} thereof --- emerges only by specifying additionally a
suitable spacetime relation on the featureless spacetime.

\begin{figure}
\centering
\includegraphics[height=6cm]{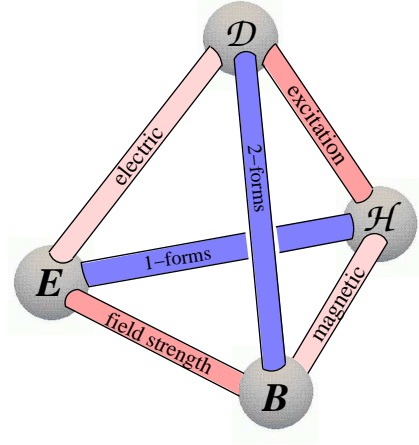}
%\picplace{5cm}{2cm} % Give the correct figure height and width in cm
%
\caption{Different aspects of the electromagnetic field. The four 
  quantities ${\cal H},{\cal D},E,B$ constitute the electromagnetic
  field. The excitations ${\cal H},{\cal D} $ are {\it extensive\/}
  quantities (how much?), the field strengths $E,B$ {\it intensive\/}
  quantities (how strong?). }
\label{fig:4}       % Give a unique label
\end{figure}

As explained, Axiom 1, Axiom 2, Axiom 3, together with Axiom 4 on
energy-momentum, constitute premetric electrodynamics. Let us display
the first three axions here again, but now Axiom 1 and Axiom 3 in in
the more general integral version. For any submanifolds $C_3$
and $C_2$ that are closed, i.e., $\partial C_3=0$ and $\partial
C_2=0$, the axioms read
\begin{equation}
\oint\limits_{C_3} J =0\,,\qquad f_\alpha=
(e_\alpha\rfloor F) \wedge J\,,\qquad
\oint\limits_{C_2}F =0\,.
\end{equation}
By de Rham's theorem we find the corresponding differential versions
\begin{eqnarray}
  d\,J&=&0\,,\qquad f_\alpha=(e_\alpha\rfloor
  F)\wedge J\,,\qquad  dF=0\,,\label{maxeqns}\\ 
  J&=&d\,H\,,\qquad\qquad\qquad\quad\qquad\qquad \hspace{3pt} F=dA\,.
\end{eqnarray}
The physical interpretation of the quantities involved  is revealed via
their $(1+3)$-decompositions \eqref{decompJ}, \eqref{HHD},
\eqref{Fdecomp}, and $ A = -\varphi d\sigma + {\cal A} $, see
Fig.\ref{fig:4}.

Let us now turn to the energy-momentum question. Using the properties
of the exterior differential, we can rewrite the Lorentz force density
\eqref{Lorentz} as
\begin{equation}
  f_\alpha = (e_\alpha\rfloor F)\wedge J =
  d\, ^ {\rm k}\! \Sigma_\alpha + X_\alpha\,.\label{fSX}
\end{equation}
Here the {\em kinematic energy-momentum} 3-form of the electromagnetic
field, a central result in the premetric electrodynamics, reads (Axiom 4)
\begin{equation}
  ^ {\rm k} \!\Sigma_\alpha :={\frac 1
    2}\left[F\wedge(e_\alpha\rfloor H) - H\wedge
      (e_\alpha\rfloor F)\right]\,.\label{simax}
\end{equation} 
The remaining force density 4-form turns out to be
\begin{equation}
X_\alpha := -{\frac 1 2}\left(F\wedge {\cal L}_{e_\alpha} H
- H\wedge{\cal L}_{e_\alpha} F \right)\,.\label{Xal}
\end{equation}
The absolute dimension of $ ^ {\rm k}\! \Sigma_\alpha$ and of
${X}_\alpha$ is $h/\ell$, where $\ell$ denotes the dimension of
length. [Provided, additionally, a linear connection is given with the
covariant differential $D$, then
\begin{equation}
  f_\alpha = D ^ {\rm k}\! \Sigma_\alpha +
  \widehat{X}_\alpha\,,\label{fSXgam}
\end{equation}
with the new supplementary force density
\begin{equation}
  \widehat{X}_\alpha = {\frac 12}\left(H\wedge{\hbox{\L}}_{e_\alpha}
    F - F\wedge{\hbox{\L}}_{e_\alpha} H\right)\,,\label{Xalgam}
\end{equation}
which contains the covariant Lie derivative. In general relativity theory,
$\widehat{X}_\alpha$ eventually vanishes for the standard Maxwell-Lorentz
electrodynamics.]

%%%%%%%%%%%%%%%%%%%%%%%%%%%%%%%%%%%%%%%%%%%%%%%%%%%%%%%%%%%%%%%%%%%
\section{The excitation is local and linear in the field strength}
%%%%%%%%%%%%%%%%%%%%%%%%%%%%%%%%%%%%%%%%%%%%%%%%%%%%%%%%%%%%%%%%%%%

The system of the Maxwell equations $dH=J\,,\,dF=0$ is apparently
under\-determined. It gets predictive power only when we supplement it
with a spacetime (or constitutive) relation between the excitation and
the field strength.  As Axiom 5, we postulate a general local and
linear spacetime relation
% with 20 (pricipal) + 15 (skewon) + 1 (axion) `constitutive' functions:
\begin{equation}\label{HchiF}
H=\kappa(F)\,,\qquad H_{ij}={\frac 1 2}\,\kappa_{ij}{}^{kl}\,F_{kl}\,.
\end{equation}
Here excitation and field strength decompose according to
$H=H_{ij}\,dx^i\wedge dx^j/2$ and $F=F_{ij}\,dx^i\wedge dx^j/2$,
respectively.  The constitutive tensor $\kappa$, as 4th rank tensor
with 36 independent components, has to be space and time dependent
since constant components would not have a generally covariant meaning
on the naked spacetime manifold we consider.

Let us decompose $\kappa_{ij}{}^{kl}$ into irreducible pieces. In the
premetric framework we can only perform a {\it contraction}. A first
contraction yields
\begin{equation}
  \kappa_i{}^k := \kappa_{il}{}^{kl}\,\quad{\rm (16\;independent\;
    functions)}\,,
\end{equation}
a second one
\begin{equation}
  \kappa := \kappa_k{}^k =
  \kappa_{kl}{}^{kl}\,\quad \hbox{(1 pseudo-scalar function)}\,.
\end{equation}
Then, introducing the traceless piece 
\begin{equation}\label{tracelesskappa}
  \not\!\kappa_i{}^k := \kappa_i{}^k - {\frac
    14}\,\kappa\,\delta_i^k\,\quad\hbox{(15 functions)}\,,
\end{equation}
we can rewrite the original constitutive tensor as
\begin{eqnarray}\nonumber
  \kappa_{ij}{}^{kl} &=& {}^{(1)}\kappa_{ij}{}^{kl} +
  {}^{(2)}\kappa_{ij}{}^{kl} + 
  {}^{(3)}\kappa_{ij}{}^{kl} \\  &=&
{}^{(1)}\kappa_{ij}{}^{kl} +
2\!\not\!\kappa_{[i}{}^{[k}\,\delta_{j]}^{l]} + {\frac
  1 6}\,\kappa\,\delta_{[i}^k\delta_{j]}^l.\label{kap-dec}
\end{eqnarray}
The {\it skewon\/} and the {\it axion\/} fields are conventionally
defined by
\begin{equation}
  \!\not\!S_i{}^j = -\,{\frac 12}\!\not\!\kappa_i{}^j,\qquad 
  \alpha = {\frac 1{12}}\,\kappa.\label{Salpha}
\end{equation}
%a traceless $4\times 4$ matrix (15 independent components) and a
%pseudo (or axial) scalar (1 component).
Substituting (\ref{kap-dec}) into (\ref{HchiF}) and using
\eqref{Salpha}, we obtain the spacetime relation explicitly:
\begin{equation}\label{crypto2a}
  H_{ij}={\frac 12}\,^{(1)}\kappa_{ij}{}^{kl}\,F_{kl}+2\,
    \!\not \!S_{[i}{}^kF_{j]k} +\alpha\,F_{ij}\,.
\end{equation} 

The {\it principal\/} (or the metric-dilaton) part
$^{(1)}\kappa_{ij}{}^{kl}$ of the constitutive tensor with 20
independent components will eventually be expressed in terms of the
metric (thereby cutting the 20 components in half). [In standard
Maxwell-Lorentz electrodynamics
\begin{equation}
 {}^{(1)}\kappa_{ij}{}^{kl} =
  \lambda_0\,\sqrt{-g}\,\hat{\epsilon}_{ijmn}\,g^{mk}g^{nl}\,,\quad
  {\!\not \!S}_{i}{}^k=0\,,\quad \alpha=0.]\label{ML}
\end{equation}
The principal part $^{(1)}\kappa_{ij}{}^{kl}$ must be non-vanishing in
order to allow for electromagnetic wave propagation in the geometrical
optics limit, see the next section. The skewon part ${\not \!S}_{i}{}^k$
with its 15 components was proposed by us. We put forward the
hypothesis that such a field exists in nature. Finally, the axion part
$\alpha$ had already been introduced in elementary particle physics in
a different connection but with the same result for electrodynamics,
see, e.g., Wilczek's axion electrodynamics \cite{Wilczek87} and the
references given there. 

The spacetime relation we are discussing here is the constitutive
relation for spacetime, i.e., for the vacuum.  However, one has
analogous structures for a {\it medium\/} described by a local and
linear constitutive law.  The {\it skewon\/} piece in this framework
corresponds to chiral properties of the medium inducing optical
activity, see Lindell et al.\ \cite{Lindell94}, whereas the concept of
an {\it axion\/} piece has been introduced by Tellegen
\cite{Tellegen1948,Tellegen1956/7} for a general medium, by
Dzyaloshinskii \cite{Dzyaloshinskii} specifically for Cr$_2$O$_3$, and
by Lindell \& Sihvola \cite{LindSihv2004a} in the form of the
so-called perfect electromagnetic conductor (PEMC). Recently, Lindell
\cite{Lindell2005} discussed the properties of a {\it
skewon-axion\/} medium.

The following alternative representation of the constitutive tensor is
useful in many derivations and for a comparison with literature, see
Post \cite{Post},
\begin{equation}
  \chi^{ijkl} := {\frac 1
    2}\,\epsilon^{ijmn}\,\kappa_{mn}{}^{kl}\,,\label{chikap}
\end{equation}
with
\begin{equation}\label{result1}
  \hspace{-20pt} \chi^{ijkl}={}
  \underbrace{^{(1)}\chi^{ijkl}}_{20,\,\text{principal}}
  +\underbrace{\epsilon^{ijm[k}\!\not\!S_m{}^{l]}
    -\epsilon^{klm[i}\!\not\!S_m{}^{j]}}_{15,\,
    \text{skewon}}+\underbrace{\epsilon^{ijkl}\,
    \alpha\;}_{1,\, \text{axion}}\,.
\end{equation}
%
%Transformation from holonomic to anholonomic coordinates with
%$e_\alpha=e^i{}_\alpha\,\partial_i$:
%$$\underbrace{\chi^{ijkl}}_{holon.}=
%e^i{}_\alpha\,e^j{}_\beta\,e^k{}_\gamma\,
%e^l{}_\delta\,\underbrace{\chi^{\alpha\beta\gamma\delta}}_{anholon.}$$

It is convenient to consider the excitation $H$ and the field strength $F$ 
as 6-vectors, each comprising a pair of two 3-vectors. The spacetime 
relation then reads
\begin{equation}
  \left(\begin{array}{c} {\cal H}_a \\ {\cal D}^a\end{array}\right) 
= \left(\begin{array}{cc} {{\mathfrak C}}^{b}{}_a & {{\mathfrak B}}_{ba} \\ 
{{\mathfrak A}}^{ba}& {{\mathfrak D}}_{b}{}^a \end{array}\right) \left(
\begin{array}{c} -E_b\\  {B}^b\end{array}\right)\,.\label{CR'}
\end{equation}
Accordingly, the constitutive tensors are represented by the $6\times
6$ matrices
\begin{equation}\label{kappachi}
  \kappa_I{}^K=\left(\begin{array}{cc} {{\mathfrak C}}^{b}{}_a & {{\mathfrak
          B}}_{ba} \\ {{\mathfrak A}}^{ba}& {{\mathfrak D}}_{b}{}^a
    \end{array}\right)\,,\qquad \chi^{IK}= \left( \begin{array}{cc}
      {\mathfrak B}_{ab}& {\mathfrak D}_a{}^b \\ {\mathfrak C}^a{}_b 
& {\mathfrak A}^{ab}\end{array}\right)\,.
\end{equation}
The $3\times 3$ matrices $\mathfrak{A,B,C,D}$ are defined by
\begin{eqnarray}\label{AB-matrix0}
{\mathfrak A}^{ba}&:=& \chi^{0a0b}\,,\qquad
{\mathfrak B}_{ba} := \frac{1}{4}\,\hat\epsilon_{acd}\,
\hat\epsilon_{bef} \,\chi^{cdef}\,,\\ \label{CD-matrix0}
{\mathfrak C}^a{}_b& :=&\frac{1}{2}\,\hat\epsilon_{bcd}\,\chi^{cd0a}\,,\qquad
{\mathfrak D}_a{}^b := \frac{1}{2}\,\hat\epsilon_{acd}
\,\chi^{0bcd}\,,
\end{eqnarray}
or explicitly, recalling the irreducible decomposition (\ref{result1}),
\begin{eqnarray}
  {\mathfrak A}^{ab} &=& - \varepsilon^{ab} -
  \epsilon^{abc}\,\!\not\!S_c{}^0,\qquad {\mathfrak B}_{ab} =
  \mu_{ab}^{-1} + \hat{\epsilon}_{abc}\,\!\not\!S_0{}^c,\label{AB}\\ 
  {\mathfrak C}^a{}_b &=&\>\;\gamma^a{}_b\, - (\!\not\!S_b{}^a 
 - \delta_b^a\,\!\not\!S_c{}^c) + \alpha\,\delta_b^a,\label{C}\\ 
  {\mathfrak D}_a{}^b &=&\>\;\gamma^b{}_a\, + (\!\not\!S_a{}^b 
 - \delta_a^b\,\!\not\!S_c{}^c) + \alpha\,\delta_a^b.\label{D}
\end{eqnarray}
The constituents of the principal part are the {\it permittivity\/}
tensor $\varepsilon^{ab}=\varepsilon^{ba}$, the {\it impermeability\/}
tensor $\mu^{-1}_{ab}= \mu_{ba}^{-1}$, and the magnetoelectric
cross-term $\gamma^a{}_b$, with $\gamma^c{}_c =0$ (Fresnel-Fizeau
effect). The skewon ${\not\!S}_b{}^a$ and the axion $\alpha$ describe
electric and magnetic Faraday effects and (in the last two relations)
optical activity. If we substitute \eqref{AB},\eqref{C},\eqref{D} into
\eqref{CR'}, we find a 3-dimensional explicit form of our Axiom 5
formulated in \eqref{HchiF}:
\begin{eqnarray}\label{explicit'}
  {\cal D}^a\!&=\!&\left( \varepsilon^{ab}\, -
    \epsilon^{abc}\not\!S_c{}^0 \right)E_b\,+\left(\hspace{9pt}
    \gamma^a{}_b +\not\! S_b{}^a - \delta_b^a\not\!  S_c{}^c\right)
  {B}^b+\alpha B^a \,,\\ {\cal H}_a\! &=\! &\left( \mu_{ab}^{-1} -
    \hat{\epsilon}_{abc}\not\!S_0{}^c \right) {B}^b +\left(-
    \gamma^b{}_a + \not\!S_a{}^b - \delta_a^b \not\!S_c{}^c\right)E_b
  - \alpha\,E_a\,.\label{explicit''}
\end{eqnarray}

%%%%%%%%%%%%%%%%%%%%%%%%%%%%%%%%%%%%%%%%%%%%%%%%%%%%%%%%%%%%%%%%%%%
\section{Propagation of electromagnetic rays (``light'')}\label{propagation}
%%%%%%%%%%%%%%%%%%%%%%%%%%%%%%%%%%%%%%%%%%%%%%%%%%%%%%%%%%%%%%%%%%%

After the spacetime relation (Axiom 5) has been formulated, we have a
complete set of equations describing the electromagnetic field. We can
now study the propagation of electromagnetic waves \`a la Hadamard.
The sourceless Maxwell equations read
\begin{equation}
dH=0\,,\quad dF=0\,.
\end{equation}
In the geometric optics approximation (equivalently, in the Hadamard
approach)
an electromagnetic wave is described by the propagation of a discontinuity 
of the electromagnetic field. The surface of discontinuity $S$ is 
defined locally by a function $\Phi$ such that $\Phi= const$ on $S$. 
The jumps $[\;]$ of the electromagnetic quantities across $S$ and the wave 
covector $q:=d\Phi$ then satisfy the geometric Hadamard conditions:
\begin{eqnarray}\label{jump}
&&[H] = 0\,,\;[dH]=q\wedge h = 0\;\qquad \Rightarrow \qquad h=q\wedge c\,,\\ 
&&\,[F] = 0\,,\;[dF]=q\wedge f = 0\;\qquad \Rightarrow \qquad f=q\wedge a\,.
\end{eqnarray}
Here $c$ and $a$ are arbitrary 1-forms. 

We use the spacetime relation and find for the jumps of the field
derivatives
\begin{equation}\label{constjump}
h=\kappa(f)=\widetilde{\kappa}(f)+\alpha f\,,
\end{equation}
with $\widetilde{\kappa}:=\,^{(1)}\kappa+{} ^{(2)}\kappa$.
Accordingly,\footnote{Compare the corresponding tensor analytical
  formula $\partial_\beta
  \widetilde{\chi}^{\alpha\beta\gamma\delta}\partial_\gamma
  A_\delta=0$ (see Post \cite{Post}, Eq.(9.40) for
  ${\chi}^{[\alpha\beta\gamma\delta ]}=0$).}
\begin{equation}
q\wedge h={q\wedge\widetilde{\kappa}(q\wedge a)=0\,}.\label{qwh}
\end{equation}
This equation is a 3-form with 4 components. We have to solve it with
respect to $a$.  As a first step, we have to remove the gauge freedom
$ a\rightarrow a+q\, \varphi$ present in (\ref{qwh}). We choose the
gauge $ \vartheta^{\hat{0}} \stackrel{*}{=}q$. After some heavy
algebra, we find (see \cite{Birkbook} for details, $a,b,...=1,2,3$)
\begin{equation}
  W^{ab}a_b \stackrel{*}{=} {0}\,,\quad{\rm with}\quad W^{ab}
  :=\widetilde{\chi} ^{\hat{0}a\hat{0}b}\,.
\end{equation}
These are 3 equations for three $a_b$'s! Nontrivial solutions exist
provided
\begin{equation}\label{fresnel1}
  {\cal W}:=\det W^{ab} \stackrel{*}{=}
  \frac{1}{3!}\hat{\epsilon}_{abc}\hat{\epsilon}_{def}
  W^{ad}W^{be}W^{cf}\stackrel{*}{=} 0.
\end{equation}
% or
% \begin{equation}\label{fresnel2}
%  {\cal W} {\stackrel{*}{=}}
%  \frac{1}{3!}\hat{\epsilon}_{abc}\hat{\epsilon}_{def}
%  \widetilde{\chi}^{\hat{0}a\hat{0}d} \widetilde{\chi}^{\hat{0}b\hat{0}e}
%  \widetilde{\chi}^{\hat{0}c\hat{0}f}{\stackrel{*}{=}} 0\,.\nonumber
% \end{equation}
We can rewrite the latter equation in a manifestly 4-dimensional
covariant form (${\hat{\epsilon}_{ a b c}\equiv \hat{\epsilon}_{\hat 0
    a b c}}$, $e_i{}^{\hat 0}\stackrel{*}{=}q_i$),
\begin{equation}\label{wgen} 
  {\cal W}=\frac{\theta^2}{4!}\,\hat{\epsilon}_{mnpq}\,
  \hat{\epsilon}_{rstu}\, {\widetilde{\chi}}{}^{\,mnri}\,
  {\widetilde{\chi}}{}^{\,jpsk}\,{\widetilde{\chi}}{}^{\,lqtu } \,
  q_iq_jq_kq_l=0\, ,\nonumber
\end{equation} 
with $\theta:=\det(e_i{}^\alpha)$. The 4-dimensional tensorial
transformation behavior is obvious.

We define 4th-order {T}amm--{R}ubilar (TR) tensor density of weight $+1$,
\begin{equation}\label{G4}
  \boxed{{\cal G}^{ijkl}(\chi):=\frac{1}{4!}\,\hat{\epsilon}_{mnpq}\,
    \hat{\epsilon}_{rstu}\, {\chi}^{mnr(i}\, {\chi}^{j|ps|k}\,
    {\chi}^{l)qtu }\,.}
\end{equation} 
It is totally symmetric ${\cal G}^{ijkl}(\chi)= {\cal G}^{(ijkl)}(\chi)$.
Thus, it has 35 independent components. Because
 $\chi^{ijkl}=\widetilde\chi^{ijkl}+\alpha\,
 \epsilon^{ijkl}$, the total antisymmetry of $\epsilon$ yields ${\cal
   G}(\chi)={\cal G}(\widetilde\chi).$ An explicit calculation shows that
\begin{equation}\label{propg8}
{\cal G}^{ijkl}(\chi) = {\cal G}^{ijkl}({}^{(1)}\chi) + {}^{(1)}
\chi^{\,m(i|n|j}\!\not\!S_m^{\ k}\!\not\!S_n^{\ l)}\,.
\end{equation}

Summarizing, we find that the wave propagation is governed by the extended 
Fresnel equation that is generally covariant in 4 dimensions:
\begin{equation} \label{Fresnel} 
 \boxed{{\cal G}^{ijkl}(\widetilde\chi)\,q_i q_j q_k q_l = 0 \,.}  
\end{equation} 
%This is the {\it quartic} equation in $q_i$ derived from a determinant 
%of a $3\times 3$ matrix.
The wave covectors $q$ lie on a {\em quartic Fresnel wave surface},
not exactly what we are observing in vacuum at the present epoch of
our universe.  Some properties of the TR-tensor, see \cite{Guillermo},
were discussed recently by Beig \cite{Beig}.

\subsection*{Extended Fresnel equation decomposed into time and space}

Recalling the `6-vector' form of the spacetime relation (\ref{CR'})
with the $3\times 3$ constitutive matrices (\ref{AB-matrix0}) and
(\ref{CD-matrix0}), we can decompose the TR-tensor into time and space
pieces: ${\cal G}^{0000} =: M\,,\ {\cal G}^{000a} =:\frac{1}{4}M^a\,,\ 
{\cal G}^{00ab} =: \frac{1}{6}M^{ab}\,,\ {\cal G}^{0abc} =:
\frac{1}{4}M^{abc}\,,\ {\cal G}^{abcd} =: M^{abcd}.$ Then the Fresnel
equation (\ref{Fresnel}) reads
\begin{equation}
  q_0^4 \underbrace{M}_{M_0} + q_0^3\underbrace{q_a\,M^a}_{M_1} +
  q_0^2\underbrace{q_a q_b\,M^{ab}}_{M_2} + q_0\underbrace{q_a q_b
    q_c\,M^{abc}}_{M_3} + \underbrace{q_a q_b q_c
    q_d\,M^{abcd}}_{M_4}=0\,,\label{fresnel}
\end{equation} 
or
\begin{equation}
M_0\,q_0^4 + M_1\,q_0^3 + M_2\,q_0^2 + M_3\,q_0 + M_4=0\,,\label{Fren}
\end{equation}
with
\begin{equation}
M =\det{\mathfrak A}\,,\qquad M^a = -\hat{\epsilon}_{bcd}\left( {\mathfrak
      A}^{ba}\,{\mathfrak A}^{ce}\, {\mathfrak C}^d_{\ e} + 
{\mathfrak A}^{ab}\,{\mathfrak
 A}^{ec}\,{\mathfrak D}_e^{\ d} \right)\,,\label{ma1}
\end{equation}
\begin{eqnarray}
  M^{ab} &=& \frac{1}{2}\,{\mathfrak A}^{(ab)}\left[({\mathfrak
      C}^d{}_d)^2 + ({\mathfrak D}_c{}^c)^2 - ({\mathfrak C}^c{}_d +
    {\mathfrak D}_d{}^c)( {\mathfrak C}^d{}_c + {\mathfrak
      D}_c{}^d)\right]\nonumber\\ && +({\mathfrak C}^d{}_c +
  {\mathfrak D}_c{}^d)({\mathfrak A}^{c(a}{ \mathfrak C}^{b)}{}_d +
  {\mathfrak D}_d{}^{(a}{\mathfrak A}^{b)c}) - {\mathfrak C}^d{}_d
  {\mathfrak A}^{c(a}{\mathfrak C}^{b)}{}_c\nonumber\\ && - {\mathfrak
    D}_c{}^{(a}{\mathfrak A}^{b)c}{\mathfrak D}_d{}^d -{\mathfrak
    A}^{dc}{\mathfrak C}^{(a}{}_c {\mathfrak D}_d{}^{b)}\nonumber\\ &&
  +\left({\mathfrak A}^{(ab)}{\mathfrak A}^{dc}- {\mathfrak
      A}^{d(a}{\mathfrak A}^{b)c}\right){\mathfrak
    B}_{dc}\,,\label{ma2}\\ M^{abc}
  &=&\epsilon^{de(c|}\left[{\mathfrak B}_{df}( {\mathfrak A}^{ab)}
    \,{\mathfrak D}_e^{\ f} - {\mathfrak D}_e^{\ a}{\mathfrak
      A}^{b)f}\,) + {\mathfrak B}_{fd}({\mathfrak A}^{ab)}\,{\mathfrak
      C}_{\ e}^f - {\mathfrak A}^{f|a}{\mathfrak C}_{\ e}^{b)})
  \right. \nonumber \\ && \left. +{\mathfrak C}^{a}_{\ f}\,{\mathfrak
      D}_e^{\ b)} \,{\mathfrak D}_d^{\ f} + {\mathfrak D}_f^{\ 
      a}\,{\mathfrak C}^{b)}_{\ e} \,{\mathfrak C}^{f}_{\ d} \right]
  \,,\label{ma3}\\ M^{abcd} &=&
  \epsilon^{ef(c}\epsilon^{|gh|d}\,{\mathfrak B}_{hf}
  \left[\frac{1}{2} \,{\mathfrak A}^{ab)}\,{\mathfrak B}_{ge} -
    {\mathfrak C}^{a}_{\ e}\,{\mathfrak D}_g^{\ b)}\right]
  \,.\label{ma4}
\end{eqnarray}

\begin{figure}
\centering
\includegraphics[height=8cm]{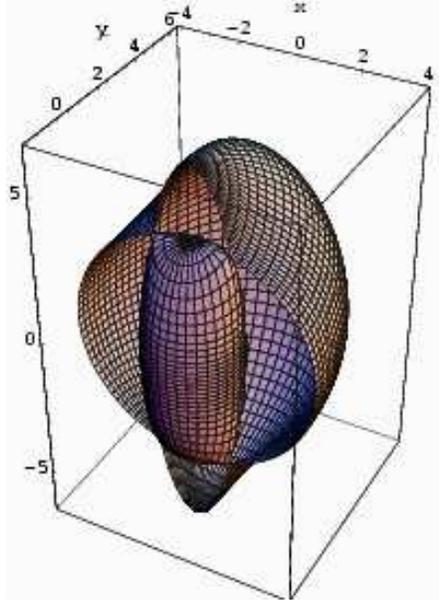}
\caption{Fresnel wave surface for anisotropic permittivity 
  $\varepsilon^{ab}= \hbox{diag}(39.7,15.4,2.3)$ and trivial
  impermeability $\mu^{-1}_{ab}=\mu_0^{-1}\hbox{diag}(1,1,1)$. The
  skewon field vanishes.  There are two branches, the outer part of
  the surface is cut into half in order to show the inner branch.  We
  use the dimensionless variables $x := cq_1/q_0,\, y := cq_2/q_0,\, z
  := cq_3/q_0$.}

% \caption{Fresnel wave covector surface in the Minkowski spacetime for an 
% anisotropic dielectric medium with $\varepsilon_1 = 39.7$, $\varepsilon_2 = 
% 15.4$, $\varepsilon_3 = 2.3$. The skewon and axion are absent. There are two 
% branches, the outer part of the surface is cut into 
% half in order to show the inner branch; we use the dimensionless 
% variables $x := cq_1/q_0,\,y := cq_2/q_0,\,z := cq_3/q_0$.}
\label{anisotropic}       % Give a unique label
\end{figure}

\subsection*{Fresnel wave surfaces}

Let us look at some Fresnel wave surfaces in order to get some feeling
for the physics involved. Divide \eqref{fresnel} by $q_0^4$ (here
$q_0$ is the frequency of the wave) and introduce the dimensionless
variables ($c$ = velocity of light in special relativity)
\begin{equation}\label{variables}
  x_a:=c\,\frac{q_a}{q_0}\,.
\end{equation}
Then we have
\begin{equation}\label{fresnelx}
  M +x_a\frac{M^a}{c} + x_ax_b\frac{M^{ab}}{c^2} + x_ax_bx_c
  \frac{M^{abc}}{c^3} + x_ax_bx_cx_d \frac{M^{abcd}}{c^4}=0\,.
\end{equation} 
We can draw these {\it quartic\/} surfaces in the dimensionless
variables $x=x_1$, $y=x_2$, $z=x_3$, provided the $M$'s are given.
According to \eqref{ma1}-\eqref{ma4}, the $M$'s can be expressed in
terms of the $3\times 3$ matrices
$\mathfrak{A},\mathfrak{B},\mathfrak{C},\mathfrak{D}$. These matrices
are specified in \eqref{AB}-\eqref{D} in terms of the permittivity
etc.. A comparison with the spacetime relations in the form of
\eqref{explicit'},\eqref{explicit''} is particularly instructive.

\begin{figure}
\centering
\includegraphics[height=7cm]{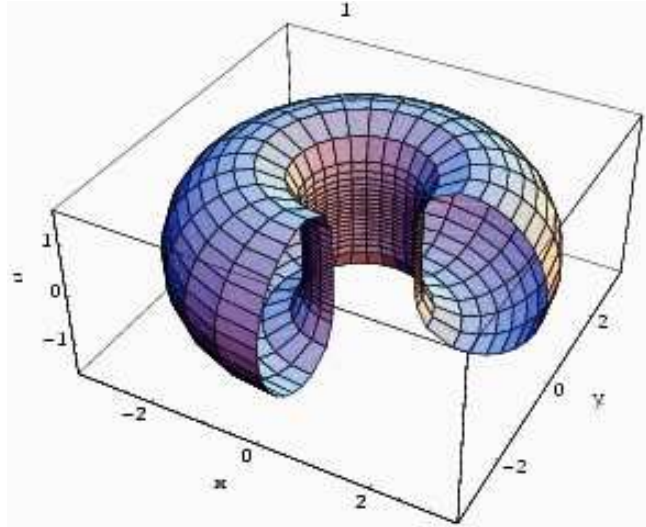}
\caption{Fresnel wave surface for trivial permittivity 
  $\varepsilon^{ab}= \varepsilon_0\,\hbox{diag}(1,1,1)$ and trivial
  impermeability $\mu^{-1}_{ab}=\mu_0^{-1}\hbox{diag}(1,1,1)$ with a
  skewon field of electric Faraday type ${\not\!S}_3{}^0=3.1\lambda_0$
  (all other components vanish). The surface has the form of a toroid
  (depicted with two cuts). We use the dimensionless variables $x :=
  cq_1/q_0,\, y := cq_2/q_0,\, z := cq_3/q_0$.}

% \caption{Fresnel surface for a real skewon field of the electric 
% Faraday type in the Minkowski spacetime. It has the form of a toroid 
% (depicted with two cuts). Here $\not\!S_i{}^j = 3.1\lambda_0\,\delta_i^3
% \delta_0^j$; we use the dimensionless variables $x :=
%   cq_1/q_0,\, y := cq_2/q_0,\, z := cq_3/q_0$.}
\label{skewonElectricFaraday}       % Give a unique label
\end{figure}
 
Let us start with a simple example. We assume that the permittivity is
anisotropic but still diagonal, $\varepsilon^{ab}=
\hbox{diag}(\varepsilon_1,\varepsilon_2,\varepsilon_3)$, whereas the
impermeability is trivial
$\mu^{-1}_{ab}=\mu_0^{-1}\hbox{diag}(1,1,1)$. No skewon field is
assumed to exist. Whether an axion field is present or not doesn't
matter since the axion does not influence the light propagation in the
geometrical optics limit. With Mathematica programs written by
Tertychniy \cite{Sergey}, we can construct for any values of
$\varepsilon_1, \varepsilon_2, \varepsilon_3$ the Fresnel wave
surface; an example is displayed in Fig.\ref{anisotropic}.

\begin{figure}
\centering
\includegraphics[height=7cm]{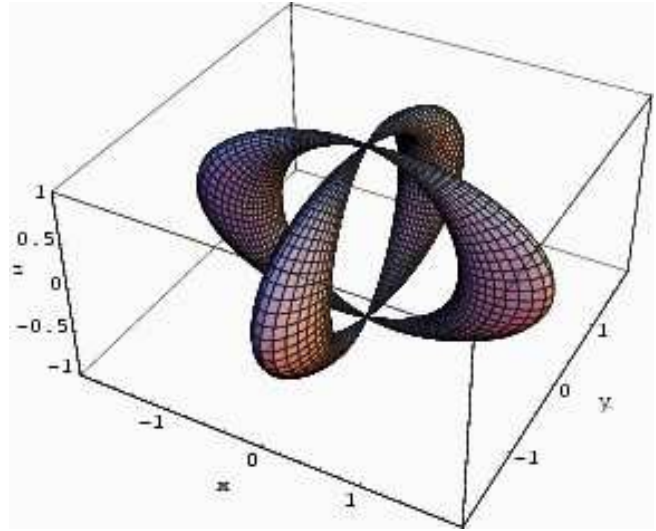}
\caption{Fresnel wave surface for trivial permittivity 
  $\varepsilon^{ab}= \varepsilon_0\,\hbox{diag}(1,1,1)$ and trivial
  impermeability $\mu^{-1}_{ab}=\mu_0^{-1}\hbox{diag}(1,1,1)$ with a
  skewon field of the magneto-electric optical activity type
  $\not\!S_1{}^2 = \not\!S_2{}^1=0.8\,\lambda_0$ (all other components
  vanish). It has two intersecting toroidal branches. We use the
  dimensionless variables $x := cq_1/q_0,\, y := cq_2/q_0,\, z :=
  cq_3/q_0$.}

% \caption{Fresnel surface for a skewon of the magneto-electric optical 
% activity type. It has two intersecting toroidal branches for a real skewon 
% $\not\!S_i{}^j = 0.8\lambda_0\left(\delta_i^1\,\delta_2^j + \delta_i^2
% \,\delta_1^j\right)$. We use the dimensionless 
% variables $x := cq_1/q_0,\,y := cq_2/q_0,\,z := cq_3/q_0$.}
\label{skewonOptical}       % Give a unique label
\end{figure}

More complicated cases are trivial permittivity $\varepsilon^{ab}=
\varepsilon_0\,\hbox{diag}(1,1,1)$ and trivial impermeability
$\mu^{-1}_{ab}=\mu_0^{-1}\hbox{diag}(1,1,1)$, but a nontrivial skewon
field. We can take a skewon field of electric Faraday type
${\not\!S}_3{}^0$, for example, see Fig.\ref{skewonElectricFaraday},
or of magnetoelectric optical activity type ${\not\!S}_1{}^2 =
{\not\!S}_2{}^1$, see Fig.\ref{skewonOptical}. In both figures and in
the subsequent one $\lambda_0 = \sqrt{\varepsilon_0/\mu_0}$ is the
admittance of free space. The characteristic feature of the skewon
field is the emergence of specific {\it holes in the Fresnel
  surfaces\/} that correspond to the directions in space along which
the wave propagation is damped out completely \cite{skewon}.  This
effect is in agreement with our earlier conclusion on the dissipative
nature of the skewon field.

\begin{figure}
\centering
\includegraphics[height=7cm]{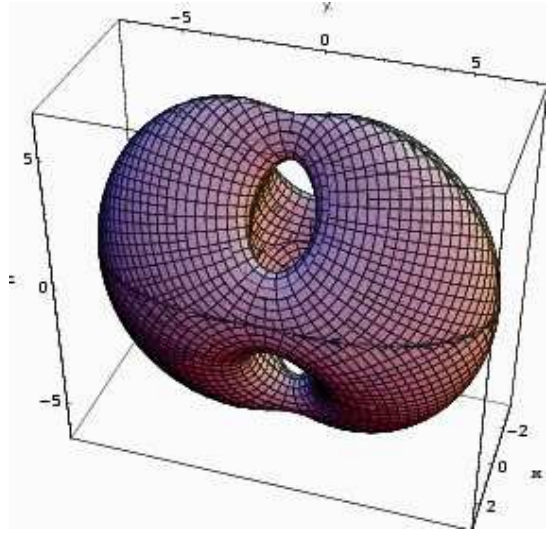}
\caption{Fresnel wave surface for anisotropic permittivity 
  $\varepsilon^{ab}= \hbox{diag}(2.4,14.8,54)$ and trivial
  impermeability $\mu^{-1}_{ab}=\mu_0^{-1}\hbox{diag}(1,1,1)$ with a
  spatially isotropic skewon field $\not\!S_1{}^1 =\not\!S_2{}^2
  =\not\!S_3{}^3 =-\frac 13\not\!S_0{}^0 = 0.25\,\lambda_0$ (all other
  components vanish). We use the dimensionless
  variables $x := cq_1/q_0,\,y := cq_2/q_0,\,z := cq_3/q_0$.}
% \caption{Fresnel covector surface for an anisotropic dielectric 
%   medium with $\varepsilon_1 = 2.4$, $\vare\not\!S =psilon_2 = 14.8$,
%   $\varepsilon_3 = 54$ and the spatially isotropic skewon $S =
%   0.5\,\lambda_0$, where $\lambda_0$ is the admittance of free space,
%   i.e., $1/(377\;\Omega).$ We use the dimensionless variables $x :=
%   cq_1/q_0,\,y := cq_2/q_0,\,z := cq_3/q_0$.}
\label{isotropicSkewon}       % Give a unique label
\end{figure}

Now we can combine anisotropic permittivity with the presence of a
skewon field. Then we expect to find some kind of
Fig.\ref{anisotropic} ``enriched'' with holes induced by the skewon
field. This time we choose a spatially isotropic skewon field with
$\not\!S_1{}^1 =\not\!S_2{}^2 =\not\!S_3{}^3 =-\frac
13\not\!S_0{}^0\ne 0$. The outcome is depicted in
Fig.\ref{isotropicSkewon}. The four holes confirm our expectation.

%%%%%%%%%%%%%%%%%%%%%%%%%%%%%%%%%%%%%%%%%%%%%%%%%%%%%%%%%%%%%%%%%%%
\section{No birefringence in vacuum and the light cone}\label{vanishing}
%%%%%%%%%%%%%%%%%%%%%%%%%%%%%%%%%%%%%%%%%%%%%%%%%%%%%%%%%%%%%%%%%%%

The propagation of light in {\it local} and {\it linear\/} premetric
vacuum electrodynamics is characterized by the extended Fresnel
equation (\ref{Fresnel}) or (\ref{Fren}). We can solve the Fresnel
equation with respect to the frequency $q_0$, keeping the 3--covector
$q_a$ fixed. With the help of Mathematica, we found the following four
solutions \cite{Laemmerzahl04}:
\begin{eqnarray}\label{freqk1+}
q_{0(\pm)}^{\uparrow} & = & \sqrt{\alpha} \pm \sqrt{\beta
+ \frac{\gamma}{\sqrt{\alpha}}} - \delta\, , \\ 
%\label{freqk2+}q_{0(2)}^{\uparrow} & = & \sqrt{\alpha} - \sqrt{\beta
%+ \frac{\gamma}{\sqrt{\alpha}}} - \delta\, ,\\ 
\label{freqk1-} q_{0(\pm)}^{\downarrow} & = & -\sqrt{\alpha} \pm\sqrt{\beta
-\frac{\gamma}{\sqrt{\alpha}}} - \delta\,.
%\\ \label{freqk2-}q_{0(2)}^{\downarrow} & = & -\sqrt{\alpha} - \sqrt{\beta
%-\frac{\gamma}{\sqrt{\alpha}}} - \delta \,.
\end{eqnarray}
We introduced the abbreviations
\begin{eqnarray}
  \alpha & := & \frac{1}{12M_0}\left(\frac{
      a}{\left(b + \sqrt{c}\right)^{\frac{1}{3}}} + \left(b +
      \sqrt{c}\right)^{\frac{1}{3}} - 2M_2\right)+\delta^2\,,\\
  \beta & := & \frac{1}{12M_0}\left(-\frac{ a}{\left(b +
        \sqrt{c}\right)^{\frac{1}{3}}} - \left(b +
      \sqrt{c}\right)^{\frac{1}{3}} -  4M_2\right)+2\delta^2\,,\\
\gamma & := & \frac{1}{4M_0}\left(2\delta M_2 - M_3\right)-2\delta^3\,,
\qquad \delta :=\frac{M_1}{4M_0}\,,
\end{eqnarray}
with
\begin{eqnarray}
a & := & 12M_0M_4 - 3M_1M_3 + M_2^2 \,,\\ 
b & := & \frac{27}{2}M_0M_3^2 - 36M_0M_2M_4 -  \frac{9}{2}M_1M_2M_3 
+ \frac{27}{2}M_1^2M_4 + M_2^3\,, \\ 
c & := & {4}\left(b^2 - a^3\right)\, .
\end{eqnarray}

\subsection*{Vanishing birefringence}

\begin{figure}
\centering
\includegraphics[height=8cm]{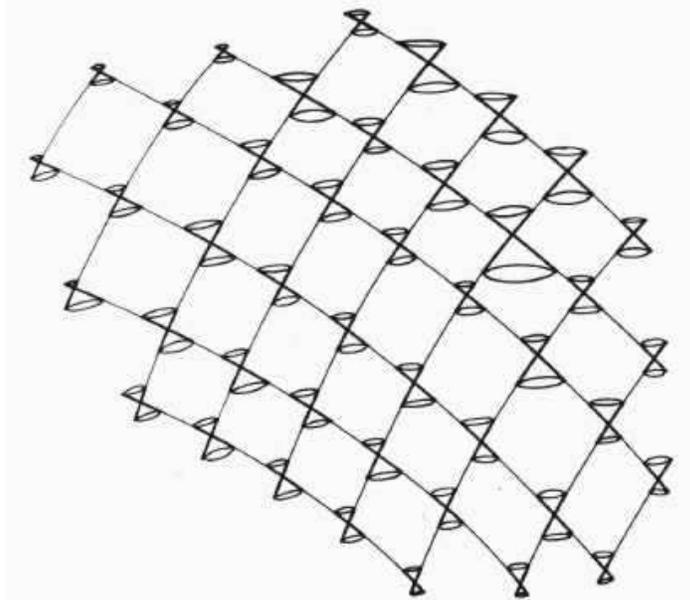}
%\picplace{5cm}{2cm} % Give the correct figure height and width in cm
%
\caption{Null cones fitted together to form a conformal manifold 
(see Pirani and Schild \cite{Pirani}).}
\label{fig:5}       % Give a unique label
\end{figure}

Now, let us demand the absence of birefringence (also called double
refraction).\footnote{Similar considerations on vanishing
  birefringence, for weak gravitational fields, are due to Ni
  \cite{Wei-Tou84}. He was also the first to understand that the axion
  field doesn't influence light propagation in the geometrical optics
  limit.} In technical terms this means, see the solutions
(\ref{freqk1+}), (\ref{freqk1-}), that $\beta =0$ and $\gamma =0$.
Then we have the degenerate solution
\begin{equation}
  q_0^\uparrow = \sqrt{\alpha} - \frac{M_1}{4M_0} \, , \qquad
  q_0^\downarrow = - \sqrt{\alpha} - \frac{M_1}{4M_0} \, .
\end{equation}
The condition $\gamma = 0$ yields directly $M_3 = M_1\left(4M_0M_2-M_1^2 
\right)/{8M_0^2}$, and, using this, we find
\begin{equation}
  \alpha = \frac{3M_1^2 - 8 M_0 M_2}{16 M_0^2}\,.
\end{equation}
Thus,
\begin{equation}
  q_0^{\uparrow\!\downarrow}=\pm\sqrt{\frac{3M_1^2-8M_0M_2}{16M_0^2}}-
  \frac{M_1}{4M_0}\,.
\end{equation}
Accordingly, the quartic wave surface (\ref{Fren}) in this case
reduces to
\begin{equation}\label{solut2}
[(q_0-q_0^{\uparrow})(q_0-q_0^{\downarrow})]^2=0\,.
\end{equation}
Multiplication yields
\begin{equation}\label{solut5}
 q_0^2+\frac 12\,\frac{M_1}{M_0}\,q_0+\frac 12\,\frac{M_2}{M_0}
 -\frac 18\,\left(\frac{M_1}{M_0}\right)^2 =0 \,.
\end{equation}
If we substitute $M_0,M_1,M_2$ as defined in (\ref{fresnel}), we have
explicitly ($i,j=0,1,2,3$)
\begin{equation}\label{solut6}
g^{ij}q_i q_j := q_0^2 + \frac 12\,\frac{M^a}{M}\, q_0q_a + \frac 18
\left(4\,\frac{M^{ab}}{M} -\frac{M^aM^b}{M^2}\right)q_aq_b = 0\,.
\end{equation}
This equation is quadratic in the 4-dimensional wave covector $q_i$.
Therefore we recover the conventional {\it light cone\/} of general
relativity at each point of spacetime, see Fig. \ref{fig:5}. Thereby
the {\it causal structure\/} of spacetime is determined. Thus, up to a
scalar factor, we derived the Riemannian metric of general relativity.

Moreover, as we have shown \cite{Birkbook,Annals}, we find the correct
{\it Lorentzian signature}. The Lorentzian (also known as Minkowskian)
signature can be traced back to the Lenz rule, which determines the
sign of the $\dot{B}$ term in the induction law.\footnote{Usually it
  is argued that the signature should be derived from quantum field
  theoretical principles; for a corresponding model, see, e.g.,
  Froggatt \& Nielsen \cite{FroNie}. Needless to say that it is our
  view that classical premetric electrodynamics together with the Lenz
  rule and a local and linear spacetime relation is all what is really
  needed.} And this sign is different from the one in the
corresponding $\dot{\cal D}$ term in the Oersted-Amp\`ere-Maxwell law.
In other words, the Lorentz signature is encoded in the decomposition
formulas \eqref{HHD} and \eqref{Fdecomp}.  Neither is the minus sign
in \eqref{HHD} a convention nor the plus sign of the $E\wedge d\sigma$
term in \eqref{Fdecomp}. Since the Lenz rule is related to the
positivity of the electromagnetic energy, the same is true for the
Lorentzian signature. This derivation of the signature of the metric
of spacetime from electrodynamics provides new insight into the
structures underlying special as well as general relativity.

%%%%%%%%%%%%%%%%%%%%%%%%%%%%%%%%%%%%%%%%%%%%%%%%%%%%%%%%%%%%%%%%%%%%
\section{Dilaton, metric, axion}
%%%%%%%%%%%%%%%%%%%%%%%%%%%%%%%%%%%%%%%%%%%%%%%%%%%%%%%%%%%%%%%%%%%%

At first the skewon and the axion emerged at the premetric level in
our theory and only subsequently the metric. Consequently, the axion
and the skewon should be regarded as more fundamental fields (if they
exist) than the metric. In the meantime, we phased out the skewon
field since we insisted, in Sec.\ref{vanishing}, on vanishing
birefringence in vacuum.

As to the metric, we recognize that multiplication of the metric by an
arbitrary function $\tilde{\lambda}(x)$ was left open in the
derivation of the last section, see \eqref{solut6}:
\begin{equation}\label{conf}
  \tilde{\lambda}(x)\,g^{ij}(x)\,q_iq_j=0\,.
\end{equation} 
Thus, only the conformally invariant part of the metric is determined.
In other words, we have actually constructed the {\it conformal\/} (or
the light cone) {\it structure} on the spacetime manifold, see, e.g.,
Weyl \cite{Weyl21,Weyl23}, Schouten \cite{Schouten54}, and Pirani \&
Schild \cite{Pirani}.

It is known from {\it special relativity\/} that the light cone (with
Lorentzian signature) is invariant under the 15-parameter conformal
group, see Barut \& R\c{a}czka \cite{Barut} and Blagojevi\'c
\cite{Milutin}. The latter, in Minkowskian coordinates $x^i$, is
generated by the following four sets of spacetime transformations:
\begin{eqnarray}
  {\rm Translations}\quad (4\ {\rm param.})\quad & x^ i &
  \rightarrow\> \tilde{x}^ i =\, x^ i +a^ i \,,\label{transl} \\ {\rm
    Lorentz\ transf.}\quad (6\ {\rm param.})\quad & x^ i &
  \rightarrow\>\tilde{x}^ i =\,\Lambda ^ i {}_ j \, x^ j
  \,,\label{lor}\\ {\rm dila(ta)tion}\quad (1\ {\rm param.})\quad &
  x^ i & \rightarrow\>\tilde{x}^ i =\,\rho\, x^ i \,,\label{dil}\\ 
  {\rm prop.\ conf.\ transf.}\quad (4\ {\rm param.})\quad & x^ i &
  \rightarrow\> \tilde{x}^i =\,{\frac{x^i + \kappa ^i\,x^2}{1+
      2\kappa_j\,x^j +\kappa_j\kappa^j\,x^2}}.\label{proper}
\end{eqnarray}
Here $a^i, \Lambda^i{}_j, \rho, \kappa^i$ are the 15 constant
parameters, and $x^2 := g_{ij}x^ix^j$. The Poincar\'e subgroup
(\ref{transl}), (\ref{lor}) (for a modern presentation of it, see
Giulini \cite{Giulini}) leaves the spacetime interval $ds^2 =
g_{ij}dx^idx^j$ invariant, whereas the dilatations (\ref{dil}) and the
proper conformal transformations (\ref{proper}) change the spacetime
interval by a scaling factor $ds^2 \rightarrow \rho^2 ds^2$ and $ds^2
\rightarrow \sigma^2 ds^2$, respectively (with $\sigma^{-1} := 1+
2\kappa_j\,x^j +\kappa_j\kappa^j\,x^2$).  In all cases the light cone
$ds^2 =0$ is left invariant. The Weyl subgroup, which is generated by
the transformations (\ref{transl})-(\ref{dil}), and its corresponding
Noether currents were discussed by, e.g., Kopczy\'nski et al.\ 
\cite{Weylcurrents}.

For massless particles, instead of the Poincar\'e group, the conformal
or the Weyl group come under consideration, since massless particles
move on the light cone.  Even though the light cone stays invariant
under all transformations \eqref{transl}-\eqref{proper}, two reference
frames that are linked to each other by a proper conformal
transformation don't stay inertial frames since their relative
velocity is not constant. If one wants to uphold the {\it inertial\/}
character of the reference frames, one has to turn to the Weyl
transformation, that is, one has to specialize to $\kappa^i=0$.

The conformal group in Minkowski space illustrates the importance of
the light cone structure on a {\it flat\/} manifold. This is
suggestive for the light cone on an arbitrarily curved manifold, even
though there is no direct relation between
\eqref{transl}-\eqref{proper} and the light cone structure we derived
in the last section.

The light cone metric $g^{ij}$ introduces the Hodge star ${}^\star$
operator.  We then can straightforwardly verify that the principal
part of the spacetime relation is determined as $H\sim{} ^\star \!F$,
where the coefficient of proportionality can be an arbitrary scalar
function $\lambda(x)$ of the spacetime coordinates. This function is
naturally identified with the {\it dilaton field}, see Brans
\cite{Carl} and Fujii \& Maeda \cite{Fujii}.  Introducing the
(Levi-Civita) dual of the excitation, $\check{H}^{ij}:=\frac
12\,\epsilon^{ijkl}\,H_{kl}$, we can then finally rewrite the
spacetime relation for vanishing birefringence in vacuum as
\begin{equation}
  \check{H}^{ij}=[\underbrace{\lambda(x)}_{\hbox{dilaton}}
  \!\!\sqrt{-g}\,g^{ik}(x)\,g^{jl}(x)+\underbrace{\alpha(x)
    }_{\hbox{axion}} \epsilon^{ijkl}\: ]\,F_{kl}\,,\label{nobirefr}
\end{equation}
that is, we are left with the constitutive fields dilaton $\lambda$,
metric $g^{ij}$, and axion $\alpha$. The combination
$\sqrt{-g}\,g^{i[k}(x)\,g^{l]j}(x)$ is conformally invariant, in
complete agreement with the above analysis.

%%%%%%%%%%%%%%%%%%%%%%%%%%%%%%%%%%%%%%%%%%%%%%%%%%%%%%%%%%%%%%%%%%%
\section{Setting the scale}
%%%%%%%%%%%%%%%%%%%%%%%%%%%%%%%%%%%%%%%%%%%%%%%%%%%%%%%%%%%%%%%%%%%

The conformal structure of spacetime is laid down in \eqref{solut6}.
Hence only 9 of the 10 independent components of the pseudo-Riemannian
metric $g_{ij}$ are specified. We need, in addition to the conformal
structure, a {\it volume measure\/} for arriving at a unique
Riemannian metric. This can be achieved by postulating a time or
length standard. 

In exterior calculus, (\ref{nobirefr}) reads
\begin{equation}
{H=\lambda(x)\,^\star \!F+ \alpha(x)\,F\,.}
\end{equation}
The axion has not been found so far, so we can put $\alpha=0$.
Moreover, under normal cicumstances, the dilaton seems to be a
constant field and thereby sets a certain scale, i.e., $\lambda(x)=
\lambda_{0}$, where $\lambda_{0}$ is the admittance of free
space\footnote{Our electrodynamical formalism is independent of the
  chosen system of units, as we discussed elsewhere \cite{Okun}.} the
value of which is, in SI-units, $1/(377\;\Omega)$.  (The exact
implementation of this assumption will have to be worked out in
future.) Accordingly, we are left with the spacetime relation of
conventional Maxwell-Lorentz electrodynamics
\begin{equation}
  \boxed{H=\lambda_0\,^\star F}\qquad \hbox{or}\qquad
  \check{H}^{ij}=\lambda_0\sqrt{-g}\,g^{ik}(x)\,g^{jl}(x)\,F_{kl}=
  \lambda_0\sqrt{-g}\,F^{ij}\,.
\end{equation}

%%%%%%%%%%%%%%%%%%%%%%%%%%%%%%%%%%%%%%%%%%%%%%%%%%%%%%%%%%%%%%%%%%%
\section{Discussion}
%%%%%%%%%%%%%%%%%%%%%%%%%%%%%%%%%%%%%%%%%%%%%%%%%%%%%%%%%%%%%%%%%%%

Weyl \cite{Weyl21,Weyl23}, in 1921, proved a theorem that the
projective and the conformal structures of a metrical space determine
its metric uniquely.  As a consequence Weyl \cite{Weyl21} argued that
{\it ...the metric of the world can be determined merely by observing
  the ``natural'' motion of material particles and the propagation of
  action, in particular that of light; measuring rods and clocks are
  not required for that.} Here we find the two elementary notions for
the determination of the metric: The paths of a freely falling point
particles, yielding the projective structure, and light rays, yielding
the conformal structure of spacetime. Later, in 1966, Pirani and
Schild \cite{Pirani}, amongst others, deepened the insight into the
conformal structure and the Weyl tensor.

In 1972, on the basis of Weyl's two primitive elements, Ehlers,
Pirani, and Schild (EPS) \cite{EPS} proposed an axiomatic framework in
which Weyl's concepts of free particles and of light rays were taken
as elementary notions that are linked to each other by plausible
axioms. Requiring compatibility between the emerging projective and
conformal structures, they ended up with a Weyl
spacetime\footnote{Time measurement in Weyl spacetime were discussed
  by Perlick \cite{Perlick91} and by Teyssandier \& Tucker
  \cite{Teyssandier95}.}(Riemannian metric with an additional
Weyl covector). They set a scale [as we did in the last section, too]
and arrived at the pseudo-Riemannian metric of general relativity. In
this sense, EPS were able to reconstruct the metric of general
relativity.

Subsequently, many authors improved and discussed the EPS-axiomatics.
Access to the corresponding literature can be found via the book of
Majer and Schmidt \cite{Majer} or the work of Perlick
\cite{Perlick91,Perlick00} and L\"ammerzahl \cite{Lammerzahl01}, e.g..
For a general review one should compare Schelb \cite{Schelb} and for a
new axiomatic scheme Schr\"oter \cite{Schroeter}.

As stated, the point particles and light rays were primary elements
that were assumed to exist and no link to mechanics nor to
electrodynamics was specified.  The particle concept within the
EPS-axiomatics lost credibility when during the emergence of gauge
theories of gravity (which started in 1956 with Utiyama \cite{Utiyama}
even before the EPS-framework had been set up in 1972) the first
quantized wave function $\Psi$ for matter entered the scene as an
elementary and ``irreducible'' concept in gravity theory. When {\it
  neutron\/} interference in an external gravitational field was
discovered experimentally in 1975 by Collella, Overhauser, and Werner
(COW) \cite{COW}, see also \cite{RauchWerner}, Sec.7, it was clear
that the point particle concept in the EPS-framework became untenable
from a physical point of view.  For completeness let us mention some
more recent experiments on matter waves in the gravitational field or
in a noninertial frame:
\begin{itemize}

\item The Werner, Staudenmann, and Colella experiment \cite{Werner79}
  in 1979 on the phase shift of neutron waves induced by the rotation
  of the Earth (Sagnac-type effect),

\item the Bonse \& Wroblewski experiment \cite{Bonse84} in 1984 on
  neutron interferometry in a noninertial frame (verifying, together
  with the COW experiment, the equivalence principle for neutron
  waves),

\item the Kasevich \& Chu interferometric experiment \cite{Kasevich}
  in 1991 with laser-cooled wave packets of sodium {\it atoms\/} in
  the gravitational field,

\item the Mewes et al.\ experiment \cite{MewesKetterle} in 1997 with
  interfering freely falling Bose-Einstein condensed sodium atoms, see
  Ketterle \cite{KetterleNobel}, Fig.14 and the corresponding text,

\item the Nesvizhevsky et al.\ experiment \cite{Nes1,Nes2} in 2002 on
  the quantum states of neutrons in the Earth's gravitational field, and

\item the Fray, H\"ansch, et al.\ \cite{Fray} experiment in 2004 with
  a matter wave interferometer based on the diffraction of atoms from
  effective absorption gratings of light. This interferometer was used
  for two stable isotopes of the rubidium atom in the gravitational
  field of the Earth. Thereby the equivalence principle was tested
  successfully on the atomic level.
\end{itemize}

Clearly, without the Schr\"odinger equation in an external
gravitational field or in a noninertial frame all these experiments
cannot be described.\footnote{A systematic procedure of deriving the
  COW result by applying the equivalence principle to the Dirac
  equation can be found in \cite{HehlNi}.} Still, in most textbooks on
gravity, these experiments are not even mentioned!

In the 1980's, as a reaction to the COW-experiment, Audretsch and
L\"ammerzahl, for a review see \cite{Audretsch}, started to develop an
axiomatic scheme for spacetime in which the point particle was
substituted by a matter wave function and the light ray be a wave
equation for electromagnetic disturbances. In this way, they could
also include projective structures with an asymmetric connection
(i.e., with torsion), which was excluded in the EPS approach a priori.

Turning to the conformal structure, which is in the center of our
interest here, L\"ammerzahl et al.\ \cite{JMP}, see also
\cite{Puntigam,HauLaem}, reconsidered the Audretsch-L\"ammerzahl
scheme and derived the inhomogeneous Maxwell equation from the
following requirements: a well-posed Cauchy problem, the superposition
principle, a finite propagation speed, and the absence of
birefringence in vacuum. The homogeneous Maxwell equation they got by
a suitable definition of the electromagnetic field strength. With a
geometric optics approximation, compare our Sec.\ref{propagation},
they recover the light ray in lowest order. And this is the message of
this type of axiomatics: Within the axiomatic system of Audretsch and
L\"ammerzahl et al., the light ray, which is elementary in the
EPS-approach, can be derived from reasonable axioms about the
propagation of electromagnetic disturbances. As with the substitution
of the mass point by a matter wave, this inquiry into the physical
nature of the light ray and the corresponding reshaping of the
EPS-scheme seems to lead to a better understanding of the metric of
spacetime. And this is exactly where our framework fits in: We also
build up the Maxwell equations in an axiomatic way and are even led to
the signature of the metric, an achievement that needs still to be
evaluated in all details.

%%%%%%%%%%%%%%%%%%%%%%%%%%%%%%%%%%%%%%%%%%%%%%%%%%%%%%%%%%%%%%%%%%%%%
\section{Summary}
%%%%%%%%%%%%%%%%%%%%%%%%%%%%%%%%%%%%%%%%%%%%%%%%%%%%%%%%%%%%%%%%%%%%%

Let us then summarize our findings: We outlined our axiomatic approach
to electrodynamics and to the derivation of the light cone. In
particular, with the help of a local and linear spacetime relation,
\begin{itemize}
\item{we found the skewon field ${\not\!S}_i{}^j$ (15 components) and
    the axion field $\alpha$ (1~component),}
\item{we found a {\it quartic\/} Fresnel wave surface for light
    propagation.}
\item{In the case of vanishing {\it birefringence,} the Fresnel wave
    surface degenerates and we recovered the light cone (determining 9
    components of the metric tensor) and, together with it, the
    conformal and causal structure of spacetime and the Hodge star
    $^\star$ operator.}
\item{If additionally the dilaton $\lambda$ (1 component) is put to a
    constant, namely to the admittance of free space $\lambda_0$
    [$1/(377\;\Omega)$ in SI-units], and the axion $\alpha$ removed,
    we recover the conventional Maxwell-Lorentz spacetime relation
    $H=\lambda_0\,^\star F$.}
\end{itemize}

\noindent Thus, in our framework, the conformal part of the {\it metric\/}
emerges from the local and linear spacetime relation as an {\em
  electromagnetic construct.} In this sense, the light cone is a
derived concept.

%%%%%%%%%%%%%%%%%%%%%%%%%%%%%%%%%%%%%%%%%%%%%%%%%%%%%%%%%%%%%%%%%%%
\section*{Acknowledgments}
%%%%%%%%%%%%%%%%%%%%%%%%%%%%%%%%%%%%%%%%%%%%%%%%%%%%%%%%%%%%%%%%%%%

One of us is grateful to Claus L\"ammerzahl and J\"urgen Ehlers for
the invitation to the Potsdam Seminar.  Many thanks go to Sergey
Tertychniy (Moscow) who provided the Mathematica programs for the
drawing of the Fresnel surfaces.  Financial support from the DFG, Bonn
(HE-528/20-1) and from {INTAS}, Brussels, is gratefully acknowledged.

%\centerline{-----------} \vspace{-12pt}
%\centerline{--------}

\printindex

\begin{thebibliography}{99}

\bibitem{Audretsch} J.~Audretsch and C.~L\"ammerzahl, {\it A new
    constructive axiomatic scheme for the geometry of space-time} In
  \cite{Majer} pp.\ 21--39 (1994).

\bibitem{Barut} A.O.\ Barut and R.\ R\c{a}czka, {\it Theory of Group
    Representations and Applications\/} (PWN -- Polish Scientific
  Publishers, Warsaw, 1977).

\bibitem{Beig} R.~Beig, {\it Concepts of Hyperbolicity and
    Relativistic Continuum Mechanics,} arXiv.org/gr-qc/0411092.

\bibitem{Milutin} M.~Blagojevi\'c, {\it Gravitation and Gauge
    Symmetries} (IOP Publishing, Bristol, 2002).

\bibitem{Bonse84} U.~Bonse and T.~Wroblewski, {\it Dynamical
    diffraction effects in noninertial neutron interferometry,}
    {\sl Phys.\ Rev.} {\bf D30} (1984) 1214--1217.  

\bibitem{Carl} C.H.~Brans, {\it The roots of scalar-tensor theory:
      An approximate history,} arXiv.org/gr-qc/0506063.
 
\bibitem{COW} R.~Colella, A.W.~Overhauser, and S.A.~Werner {\it
    Observation of Gravitationally Induced Quantum Interference}, {\sl
    Phys.~Rev.~Lett.} {\bf 34} (1975) 1472--1474.

\bibitem{Delphenich1} D.H.~Delphenich, {\it On the axioms of
    topological electromagnetism,} {\sl Ann.\ Phys. (Leipzig)} {\bf
    14} (2005) 347--377; updated version of arXiv.org/hep-th/0311256.

\bibitem{Delphenich2} D.H.~Delphenich, {\it Symmetries and pre-metric
    electromagnetism,} {\sl Ann.\ Phys. (Leipzig)} {\bf 14} (2005)
  issue 11 or 12, to be published.

% \bibitem{Dombrowski} P.~Dombrowski, {\it Wege in euklidischen Ebenen,
%     Kinematik der Speziellen Relativit\"atstheorie.} Springer Berlin
%   (1999).
% 
\bibitem{Dzyaloshinskii} I.E.~Dzyaloshinskii, {\it On the
    magneto-electrical effect in antiferromagnets,} {\sl J.\ Exptl.\ 
    Theoret.\ Phys. (USSR)} {\bf 37} (1959) 881--882 [English transl.:
    {\sl Sov.\ Phys.\ JETP} {\bf 10} (1960) 628--629].

  \bibitem{edelen} D.G.B.~Edelen, {\it A metric free electrodynamics
      with electric and magnetic charges,} {\sl Ann.\ Phys.\ (NY)}
    {\bf 112} (1978) 366--400.

\bibitem{EPS} J.~Ehlers, F.A.E.~Pirani, and A.~Schild, {\it The
    geometry of free fall and light propagation,} in: {\sl General
    Relativity, papers in honour of J.L.~Synge}, L.~O'Raifeartaigh,
  ed.\ (Clarendon Press, Oxford, 1972), pp.\ 63--84.

\bibitem{Einstein05} A.~Einstein, {\it Zur Elektrodynamik bewegter
    K\"orper,} {\sl Ann.\ Phys. (Leipzig)} {\bf 17} (1905) 891--921.
    English translation in \cite{Dover1952}.

% \bibitem{Einstein55} A. Einstein: {\it The Meaning of Relativity,} 5th
%   ed.\ (Princeton University Press, Princeton 1955).
% 
\bibitem{Fray} S.~Fray, C.~Alvarez Diez, T.~W.~H\"ansch and
  M.~Weitz, {\it Atomic interferometer with amplitude gratings of
    light and its applications to atom based tests of the equivalence
    principle,} {\sl Phys.\ Rev.\ Lett.} {\bf 93} (2004) 240404 (4
  pages); arXiv.org/physics/0411052.

\bibitem{FroNie} C.D.~Froggatt and H.B.~Nielsen, {\it Derivation of
    Poincar\'e invariance from general quantum field theory,} {\sl
    Ann.\ Phys.\ (Leipzig)} {\bf 14} (2005) 115--147 [Special
  issue commemorating Albert Einstein].

\bibitem{Fujii} Y.~Fujii and K.-I.~Maeda, {\it The Scalar-Tensor
    Theory of Gravitation} (Cambridge University Press, Cambridge,
  2003).

\bibitem{Giulini} D.~Giulini, {\it The Poincar\'e group: Algebraic,
    representation-theoretic, and geometric aspects,} these
  Proceedings (2005).

\bibitem{HauLaem} M.~Haugan and C.~L\"ammerzahl, {\it On the
    experimental foundations of the Maxwell equations}, {\sl Ann.
    Phys.\ (Leipzig)} {\bf 9} (2000) Special Issue, SI-119--SI-124.

\bibitem{HehlNi} F.W. Hehl and W.-T. Ni: {\it Inertial effects of a
    Dirac particle.} {\sl Phys.\ Rev.} {\bf D42} (1990) 2045--2048.

\bibitem{Birkbook} F.W.~Hehl and Yu.N.~Obukhov, {\em Foundations of
    Classical Electrodynamics --- Charge, Flux, and Metric}
  (Birkh{\"a}user, Boston, MA, 2003).

\bibitem{ourmonopole} F.W.~Hehl and Yu.N.~Obukhov, {\it
    Electric/magnetic reciprocity in premetric electrodynamics with
    and without magnetic charge, and the complex electromagnetic
    field,} {\sl Phys.\ Lett.} {\bf A323} (2004) 169--175;
  arXiv.org/physics/0401083.

\bibitem{Okun} F.W.~Hehl and Yu.~N.~Obukhov, {\it Dimensions and units
    in electrodynamics,} {\sl Gen.\ Rel.\ Grav.} {\bf 37} (2005)
  733--749; arXiv.org/physics/0407022.

  \bibitem{PostCon} F.W.~Hehl and Yu.N.~Obukhov, {\it Linear media in
      classical electrodynamics and the Post constraint,} {\sl Phys.\ 
      Lett.} {\bf A334} (2005) 249--259; arXiv.org/physics/ 0411038.

\bibitem{Itin} Y.~Itin, {\it Caroll-Field-Jackiw electrodynamics in
    the pre-metric framework,} {\sl Phys.\ Rev.} {\bf D70} (2004)
  025012 (6 pages); arXiv/org/hep-th/0403023.

\bibitem{Annals} Y.~Itin and F.W.~Hehl, {\it Is the Lorentz signature
    of the metric of spacetime electromagnetic in origin?} {\it Ann.\ 
    Phys.\ (NY)} {\bf 312} (2004) 60--83; arXiv.org/gr-qc/0401016.

\bibitem{Kaiser} G.~Kaiser, {\it Energy-momentum conservation in
    pre-metric electrodynamics with magnetic charges}, {\sl J.\ Phys.}
  {\bf A37} (2004) 7163--7168.

\bibitem{Kasevich} M.~Kasevich and S.~Chu, {\it Atomic interferometry
    using stimulated Raman transitions,} {\sl Phys.\ Rev.\ Lett.} {\bf
    67} (1991) 181--184.

\bibitem{KetterleNobel} W.~Ketterle, {\it When atoms behave as waves:
    Bose-Einstein condensation and the atom laser\/} (Noble lecture
  2001), 

{\tt  http://nobelprize.org/physics/laureates/2001/ketterle-lecture.pdf} .

\bibitem{Kiehn04} R.M.~Kiehn, {\it Plasmas and Non Equilibrium
    Electrodynamics} 2005 (312 pages), see {\tt
    http://www22.pair.com/csdc/download/plasmas85h.pdf} .

\bibitem{Weylcurrents} W.~Kopczy\'nski, J.D.~McCrea and F.W. Hehl,
  {\it The Weyl group and its current}. {\sl Phys.\ Lett.} {\bf A128}
  (1988) 313--317.

\bibitem{Lammerzahl01} C.~L\"ammerzahl, {\it A characterisation of the
    Weylian structure of space-time by means of low velocity tests,}
  {\sl Gen.\ Rel.\ Grav.} {\bf 33} (2001) 815--831;
  arXiv.org/gr-qc/0103047.
 
\bibitem{JMP} C.~L\"ammerzahl, A.~Camacho, and A.~Mac\'{\i}as, {\it
    Reasons for the electromagnetic field to obey Maxwell's
    equations,} submitted to {\sl J.\ Math.\ Phys.} (2005).

\bibitem{Laemmerzahl04} C.~L\"ammerzahl and F.W.~Hehl, {\it Riemannian
    light cone from vanishing birefringence in premetric vacuum
    electrodynamics}, {\sl Phys.\ Rev.} {\bf D70} (2004) 105022 (7
  pages); arXiv.org/gr-qc/0409072.

\bibitem{chargenoncons} C.~L\"ammerzahl, A.~Mac\'{\i}as, and
  H.~M\"uller, {\it Lorentz invariance violation and charge
    (non)conservation: A general theoretical frame for extensions of
    the Maxwell equations,} {\sl Phys.\ Rev.} {\bf D71} (2005) 025007
    (15 pages).

\bibitem{Lindell} I.V.~Lindell, {\it Differential Forms in
      Electromagnetics} (IEEE Press, Piscataway, NJ, and
    Wiley-Interscience, 2004).

\bibitem{Lindell2005} L.V.~Lindell, {\it The class of bi-anisotropic
      IB-media}, {\sl J.\ Electromag.\ Waves Appl.,} to be published
    (2005).

\bibitem{LindSihv2004a} I.V.~Lindell and A.H.~Sihvola, {\it Perfect
    electromagnetic conductor,} {\sl J.\ Electromag.\ Waves Appl.}
  {\bf 19} (2005) 861--869; arXiv.org/physics/0503232.

\bibitem{Lindell94} I.V.~Lindell, A.H.~Sihvola, S.A.~Tretyakov,
  A.J.~Viitanen, {\it Electromagnetic Waves in Chiral and Bi-Isotropic
    Media} (Artech House, Boston, MA, 1994).

\bibitem{Dover1952} H.A.~Lorentz, A.~Einstein, H.~Minkowski, and
  H.~Weyl, {\em The Principle of Relativity.} A collection of original
  memoirs. Translation from the German (Dover, New York, 1952).

\bibitem{Majer} U.~Majer and H.-J.Schmidt (eds.), {\it Semantical
    Aspects of Spacetime Theories} (BI-Wissenschaftsverlag, Mannheim
  1994).

\bibitem{MewesKetterle} M.-O.~Mewes et al., {\it Output Coupler for
    Bose-Einstein Condensed Atoms,} {\sl Phys.\ Rev.\ Lett.} {\bf 78}
  (1997) 582--585.

\bibitem{Nes1} V.V.~Nesvizhevsky {et al.}, {\it Quantum states of
    neutrons in the Earth's gravitational field,} {\sl Nature} {\bf
    415} (2002) 297--299.

\bibitem{Nes2} V.V.~Nesvizhevsky {et al.}, {\it Measurement of quantum
    states of neutrons in the Earth's gravitational field,} {\sl
    Phys.\ Rev.} {\bf D67} (2003) 102002 (9 pages);
  arXiv.org/hep-ph/0306198.

\bibitem{Wei-Tou84} W.-T.~Ni, {\it Equivalence principles and
    precision experiments.} In {\sl Precision Measurement and
    Fundamental Constants II,} B.N.~Taylor, W.D.~Phillips, eds.  Nat.\ 
  Bur.\ Stand.\ (US) Spec.\ Publ.\ 617 (US Government Printing Office,
  Washington, DC, 1984) pp.\ 647--651.

\bibitem{skewon} Yu.N.~Obukhov and F.W.~Hehl, {\it Possible skewon
    effects on light propagation,} {\sl Phys.\ Rev.} {\bf D70} (2004)
  125015 (14 pages); arXiv.org/physics/0409155.

\bibitem{measure} Yu.N.~Obukhov and F.W.~Hehl, {\it Measuring a
    piecewise constant axion field in classical electrodynamics,} {\sl
    Phys.\ Lett.} {\bf A341} (2005) 357--365;
  arXiv.org/ physics/0504172.

\bibitem{Perlick91} V.~Perlick, {\it Observer fields in Weylian
    spacetime models,} {\sl Class.\ Quantum Grav.} {\bf 8} (1991)
  1369--1385.

\bibitem{Perlick00} V.~Perlick, {\it Ray optics, Fermat's principle,
    and applications to general relativity,} Lecture Notes in Physics
  (Springer) {\bf m61} (2000) 220 pages.

\bibitem{Pirani} F.A.E.\ Pirani and A.\ Schild, {\it Conformal
    geometry and the interpretation of the Weyl tensor}, in: {\sl
    Perspectives in Geometry and Relativity.} Essays in honor of V.\ 
  Hlavat\'y. B.\ Hoffmann, editor (Indiana University Press,
  Bloomington, 1966) pp.\ 291--309.

\bibitem{Post} E.J.\ Post, {\it Formal Structure of Electromagnetics
    -- General Covariance and Electromagnetics} (North Holland,
  Amsterdam, 1962, and Dover, Mineola, New York, 1997).

\bibitem{Puntigam} R.A.~Puntigam, C.~L\"ammerzahl and F.W.~Hehl, {\it
    Maxwell's theory on a post-Riemannian spacetime and the
    equivalence principle,} {\sl Class.\ Quant.\ Grav.} {\bf 14}
  (1997) 1347--1356; arXiv.org/gr-qc/9607023.

\bibitem{RauchWerner} H.~Rauch and S.A.~Werner, {\it Neutron
    Interferometry,} Lessons in experimental quantum mechanics
  (Clarendon Press, Oxford, 2000).

\bibitem{Guillermo} G.F.\ Rubilar, {\it Linear pre-metric
    electrodynamics and deduction of the lightcone\/}, {\sl Thesis}
  (University of Cologne, June 2002); see {\sl Ann.\ Phys.\ (Leipzig)}
  {\bf 11} (2002) 717--782.

% \bibitem{Schaefer32} C.~Schaefer, {\it Einf\"uhrung in die
%    theoretische Physik: Dritter Band, erster Teil --- Elektrodynamik
%    und Optik}, de Gruyter, Berlin (1932).
%
\bibitem{Schelb} U.~Schelb, {\it Zur physikalischen Begr\"undung der
    Raum-Zeit-Geometrie,} Habilitation thesis (Univ.\ Paderborn, 1997).

\bibitem{Schouten54} J.A.~Schouten, {\it Ricci-Calculus}, 2nd ed.
  (Springer, Berlin, 1954).

\bibitem{Schroeter} J.~Schr\"oter, {\it A new formulation of general
    relativity.} Part I. Pre-radar charts as generating functions for
  metric and velocity. Part II. Pre-radar charts as generating
  functions in arbitrary space-times. Part III. GTR as scalar field
  theory (altogether 58 pages). Preprint, Univ.\ Paderborn (July
  2005).

\bibitem{Tellegen1948} B.D.H.~Tellegen, {\it The gyrator, a new
    electric network element,} {\sl Philips Res.\ Rep.} {\bf 3} (1948)
  81--101.

\bibitem{Tellegen1956/7} B.D.H.~Tellegen, {\it The gyrator, an
    electric network element,} in: {\it Philips Technical Review} {\bf
    18} (1956/57) 120--124. Reprinted in H.B.G.~Casimir and
  S.~Gradstein (eds.) {\it An Anthology of Philips Research.}
  (Philips' Gloeilampenfabrieken, Eindhoven, 1966) pp.\ 186--190.

\bibitem{Sergey} S.I.~Tertychniy [National Research Institute for
  Physical, Technical, and Radio-Technical Measurements (VNIIFTRI),
  141570 Mendeleevo, Russia. Email: {\tt bpt97@mendeleevo.ru}]
  provided the Mathematica programs for constructing the figures of
  the quartic Fresnel surface.

\bibitem{Teyssandier95} P.~Teyssandier and R.~W.~Tucker, {\it Gravity,
    Gauges and Clocks,} {\sl Class.\ Quant.\ Grav.} {\bf 13} (1996)
  145--152.
 
\bibitem{Utiyama} R.~Utiyama, {\it Invariant theoretical
    interpretation of interaction,} {\sl Phys.\ Rev.} {\bf 101} (1956)
  1597--1607.

\bibitem{Werner79} S.A.~Werner, J.-L.~Staudenmann, and R.~Colella,
  {\it Effect of Earth's rotation on the quantum mechanical phase of
    the neutron,} {\sl Phys.\ Rev.\ Lett.} {\bf 42} (1979) 1103--1106.

\bibitem{Weyl21} H.~Weyl, {\it Zur Infinitesimalgeometrie: Einordnung
    der projektiven und konformen Auffassung,} {\sl Nachr.\ K\"onigl.\ 
    Gesellschaft Wiss.\ G\"ottingen, Math.-Phys. Klasse,} pp.\ 99--112
  (1921); also in K.~Chandrasekharan (ed.), {\it Hermann Weyl,
    Gesammelte Abhandlungen} Vol.II, 195--207 (Springer, Berlin,
  1968).

\bibitem{Weyl23} H.~Weyl, {\it Raum, Zeit, Materie}, Vorlesungen
  \"uber Allgemeine Rela\-tivi\-t\"ats\-theorie, re\-print of the 5th
  ed.\ of 1923 (Wissenschaftliche Buchges., Darmstadt, 1961).  Engl.\ 
  translation of the 4th ed.: {\it Space--Time--Matter} (Dover Publ.,
  New York, 1952).

\bibitem{Wilczek87} F.\ Wilczek, {\it Two applications of axion
    electrodynamics}, {\sl Phys.\ Rev.\ Lett.} {\bf 58} (1987)
    1799--1802.


\end{thebibliography}
\end{document}